\documentclass[10pt]{article}
\usepackage{natbib}
\usepackage[dvips]{graphicx}
\usepackage{wrapfig,epsfig,epsf}
\usepackage{amsmath,amsthm,amsfonts,amssymb}
\usepackage{cmap}
\usepackage{natbib}
\usepackage{times}
\usepackage{aas_macros}

\usepackage[english]{babel}  
\defcitealias{mukhopadhyay-2005}{MAN05}
\defcitealias{zhuravlev-razdoburdin-2014}{ZR14}
\defcitealias{razdoburdin-zhuravlev-2015}{RZ15}

\begin{document}

\title{Transient growth of perturbations on scales beyond the accretion disc thickness}
\author{ D. N. Razdoburdin$^{1,2}$ and V. V. Zhuravlev$^{1}$\thanks{E-mail: zhuravlev@sai.msu.ru}}
\date{$^{1}$Sternberg Astronomical Institute, Moscow M.V. Lomonosov State University, Universitetskij pr., 13, Moscow 119992, Russia\\
$^{2}$Department of Physics, Moscow M.V. Lomonosov State University, Moscow, 119992, Russia}

\maketitle

\begin{abstract}

Turbulent state of spectrally stable shear flows may be developed and sustained according to the bypass scenario of transition. 
If it works in non-magnetised boundless and homogeneous quasi-Keplerian flow, transiently growing shearing vortices should supply turbulence with energy.
Employing the large shearing box approximation, as well as a set of global disc models, we study the optimal growth of the shearing 
vortices in such a flow in the whole range of azimuthal length-scales, $\lambda_y$, as compared to the flow scale-height, $H$. It is shown that 
with the account of the viscosity the highest possible amplification of shearing vortices, $G_{max}$, attains maximum at $\lambda_y\lesssim H$ 
and declines towards both the large scales $\lambda_y\gg H$ and the small scales $\lambda_y\ll H$ in a good agreement 
with analytical estimations based on balanced solutions. We pay main attention to the large-scale vortices $\lambda_y\gg H$, which 
produce $G_{max}\propto (\Omega/\kappa)^4$, where $\Omega$ and $\kappa$ denote local rotational and epicyclic frequencies, respectively.
It is demonstrated that the large-scale vortices acquire high density perturbation as they approach the instant of swing. At the same time,
their growth is not affected by bulk viscosity. We check that $G_{max}$ obtained globally is comparable to its local counterpart
and the shape and localisation of global optimal vortices can be explained in terms of the local approach. 
The obtained results allow us to suggest that the critical Reynolds number of subcritical transition to turbulence in quasi-Keplerian flow, as well as
the corresponding turbulent effective azimuthal stress should substantially depend on shear rate. 

\end{abstract}


\section{Introduction}

Homogeneous quasi-Keplerian shear flow, which is defined as a rotating flow with its specific angular momentum and angular velocity, respectively, increasing and decreasing outwards, has proved to be the most enigmatic case 
with regards to the transition to turbulence. To date, its hydrodynamic (HD) stability with respect to finite amplitude perturbations has been experimentally checked up to the Reynolds number equal to a few millions, see \citet{edlund-ji-2014}.
This is a pretty high value bearing in mind that the Reynolds number corresponding to the subcritical transition $R_T\sim 350$ for plane Couette flow and $R_T\sim 1000$ for plane Poiseille flow, see
\citet{schmid-henningson-2001} as well as numerous references therein, and $R_T\sim 7000$ for the narrow gap rotating Taylor-Couette flow with angular velocity increasing towards the periphery, see \citet{burin-2012}.
However, there is a number of well-known observational evidences that cold astrophysical discs, where magnetorotational instability ceases to operate, are accreting. At the same time, the smallness of microscopic viscosity entails huge Reynolds numbers up to $\sim10^{13}$ attained in astrophysical discs. 
Thereby, the possibility of a purely HD route to turbulence in quasi-Keplerian flow is not ruled out and still represents an important unresolved problem.

The transition to turbulence and the replenishment of turbulent energy budget both cannot take place without growth of perturbations, i.e. increase of their amplitudes due to the energy transfer from the background shear.
At the same time, it turns out that instantaneous growth rate is independent of perturbations amplitude \citep{henningson-reddy-1994}. 
This basic fact is obtained from the Reynolds-Orr equation for evolution of kinetic energy of perturbations obeying Navier-Stockes equations. 
Thus, the energy transfer from the background motion to turbulent fluctuations is given by mechanisms which are present in linear dynamical equations. 
For a subcritical transition to turbulence, which is the case also for a spectrally stable homogeneous quasi-Keplerian flow, this process has been attributed to a transient growth of perturbations which remains to be the only way to draw energy 
from the background motion in the absence of exponentially growing modes of perturbations \citep{trefethen-1993, schmid-2007}.

As it has become clear in the past decades, there are two different kinds of transiently growing perturbations and two respective variants of growth mechanisms in homogeneous shear flows. 
The first one is usually referred to as streamwise rolls \citep{butler-farrell-1992}, which generate growing streamwise streaks, i.e. the domains of the enhanced streamwise velocity, by the so called lift-up effect \citep{ellingsen-palm-1975}. 
The lift-up effect plays a key role in self-sustaining process (SSP), which provides a turbulent state of plane shear flows, see \citet{waleffe-1997}.
The bypass transition via the transient growth of streamwise rolls has been investigated well in literature (see, e.g., \citet{reddy-1998} and its citations): thanks to
a possibility to simulate the breakdown to turbulence
in plane shear flows at moderate Reynolds numbers. Recently, \citet{mamatsashvili-2016} has studied turbulence dynamics by means of the three-dimensional (3D) Fourier representation
enabling one to describe the SSP in terms of the non-linear transverse cascade, which is qualitatively different from previously known direct (or inverse) turbulence cascade.  
Yet, the situation becomes much less clear as we proceed to spectrally stable rotating shear flows. 
Basically, the Coriolis force coming into play stabilises the rolls, i.e. their transformation to streaks turns into epicyclic motion of fluid particles, which prevents transient growth. 
In such a way, another type of transiently growing perturbations comes out of the shadow: shearing vortices. 

Shearing vortices are streamwise elongated but streamwise-dependent structures enhanced via contraction by the shear provided that 
vorticity is conserved. The latter is usually called the Orr mechanism or, alternatively, the swing amplification process. In order to be amplified in the rotating flow, 
shearing vortices must be in the form of leading (trailing) spirals with respect to the background flow of quasi-Keplerian (cyclonic) type (look, e.g., \citet{lesur-longaretti-2005} for definition of rotating flow types). 
Thus, only certain spatial Fourier harmonics (SFHs) of shearing vortices can supply turbulence with energy.
Note that shearing vortices are also incorporated in bypass transition of two-dimensional (2D) models of plane shear flows, where the rolls are absent.
Another picture of the non-linear feedback sustaining turbulence should be suggested in this case, see \citet{chagelishvili-2003}. 
The novel ingredient here is a coupling of SFHs from certain quadrants of wavevector space, which repopulates the quadrant with linearly growing shearing vortices. 
\citet{lithwick-2007} has considered a particular case of such coupling between axisymmetric and shearing vortices in 2D model of Keplerian flow and showed that it produces a new leading spiral with a larger amplitude than the existing one.
In light of the present study, it is important to note that this kind of positive non-linear feedback is provided by the shearing vortices of a sufficiently large azimuthal length-scale, $\lambda_y$. 
Moreover, in later work \citet{lithwick-2009} has found that non-linear vortices emerging by this 2D mechanism survive in 3D model provided that their azimuthal length-scales are larger than the flow scale-height, $\lambda_y > H$.  
\citet{umurhan-regev-2004} and \citet{horton-2010} also studied the non-linear 2D dynamics of rotating (and plane) shear flows providing additional arguments in favour of non-linear sustenance of linearly growing shearing vortices. 
However, presently there is no theoretical understanding, whether the reservoir of linearly growing shearing vortices may be repopulated in full 3D dynamics of perturbations with not necessarily $\lambda_y > H$. 
Nevertheless, \citet{meseguer-2002} has revealed a strong correlation between a value of maximum transient growth, $G_{max}$, and a point of relaminarisation of the Rayleigh-stable incompressible flow between counter-rotating cylinders.
Apparently, this is an argument in favour of bypass transition via the swing amplification of perturbations, since $G_{max}$ in this particular regime of Taylor-Couette flow is produced by shearing vortices 
rather than by streamwise rolls. 

The quantitative characteristic of transient growth, $G_{max}$, is obtained employing the method of optimisation for solving the problem of perturbation dynamics. \citet{yecko-2004} and \citet{mukhopadhyay-2005} (hereafter \citetalias{mukhopadhyay-2005}) have also identified the shearing vortices, as well as corresponding $G_{max}$, applying this method to the dynamics of 3D incompressible HD perturbations. They have considered the problem in the shearing box approximation invented earlier by \citet{goldreich-lynden-bell-1965} in application to swing amplification in perturbed stellar discs. 
It has been confirmed that in case of the Keplerian shear the optimal growth is attained by quasi-2D columnar leading spirals exposed to swing amplification.
Additionally, \citetalias{mukhopadhyay-2005} have considered super-Keplerian shear rates, which are intermediate between the Keplerian one and the one with constant specific angular momentum distribution. 
They have found that exactly constant angular momentum shear provides the optimal vortices in the form of streamwise streaks, which are deformed by the background flow and give birth to growing streamwise rolls. This mechanism of non-modal
dynamics is exactly reverse to the lift-up effect in plane shear flows, consequently, it was termed as the 'anti lift-up effect' \citep{antkowiak-brancher-2007}. The corresponding maximum optimal growth factor is several orders of magnitude
larger than those for the Keplerian shear. Another important result is that the maximum optimal growth suffers exceedingly rapid drop as one proceeds to the shear flow with a non-zero (positive) angular momentum radial derivative, see Fig. 4 of \citetalias{mukhopadhyay-2005}. Physically, this is caused by the fact that the growth of rolls degenerates into epicyclic motions, whereas the growth of small-scale shearing vortices is much lower and {\it weakly depends on the shear rate}.
Further, the recent results of \citet{maretzke-2014} confirm that in all spectrally stable regimes of Taylor-Couette flow, rather than only in quasi-Keplerian one, the incompressible 3D perturbations attain 
the highest growth factors by virtue of swinging shearing vortices. However, though in each regime they find growth factors of the same order of magnitude, there are qualitative features of cyclonic and counter-rotating flows making them different 
from quasi-Keplerian one. Unlike in quasi-Keplerian regime, in the cyclonic and counter-rotating regimes those shearing vortices have (though weak) axial dependence and up to 80\% of kinetic energy transferred into 
the axial velocity perturbation near the transient growth maximum. This features go beyond swing amplification mechanism and \citet{maretzke-2014} note that the reason for such qualitative difference between quasi-Keplerian 
and other regimes remains unclear. Thereby, two-dimensionality of shearing vortices attaining maximum transient growth is a unique peculiarity of the quasi-Keplerian flow, which may prove to be the reason for its extraordinary non-linear HD stability.

The studies mentioned above have been dealing with incompressible perturbation dynamics, which in the context of accretion disc physics implies that perturbations have characteristic length-scale much smaller than the disc thickness.
We will refer to such perturbations as 'small-scale vortices' below in the text. In contrast, we are going to consider how a finite disc thickness affects shearing vortices in the quasi-Keplerian flow. 
\citet{zhuravlev-razdoburdin-2014} (hereafter \citetalias{zhuravlev-razdoburdin-2014}) showed analytically that the growth factors of the vortical leading spirals with azimuthal wavelength larger than the disc thickness, which we will refer to as
'large-scale vortices' below in the text, should strongly depend on the shear rate as one approaches constant angular momentum rotation, and $G_{max}\propto \kappa^{-4}$, where $\kappa$ is epicyclic frequency expressed in units 
of rotation frequency. In this work we are going to check numerically this analytical result employing a variational technique for determination of linear optimal perturbations in the shearing box approximation. 
Our task is to take into account the dynamical influence of viscosity. Besides, we also consider the influence of bulk viscosity, what is especially important to do as the spiral swings from leading to trailing one, 
since as this happens, the radial wavenumber of SFH tends to zero and one cannot distinguish between vortices and density waves anymore. Thereby, the field of the velocity perturbation acquires significant divergence, which may cause 
a suppression of transient growth of initially vortical SFH. Also, as far as the azimuthal wavenumber of SFH is of order of the inversed disc thickness, the vortical SFH generates strong density wave at the instant of swing
\citep{chagelishvili-1997,bodo-2005,heinemann-papaloizou-2009a}, whilst, the acoustic motion of the fluid is primarily affected by a finite time of the establishment of thermodynamic equilibrium, characterised by the bulk viscosity, 
see e.g. \citet{landau-lifshitz-1987}. 

Whereas in this work such kind of problem is considered for the first time in application to rotating and, particularly, quasi-Keplerian flows, \citet{hanifi-1996}, \citet{farrell-ioannou-2000} and \citet{malik-2006} 
have already studied it with an interest to compressible plane shear flows. They found that the lift-up effect remains to be valid with the account of finite Mach number, and that the streamwise rolls remain to be the
optimal perturbations in the flow. With regards to the shearing vortices in compressible 
medium, \citet{farrell-ioannou-2000} have found their growth factors being considerably larger than corresponding incompressible values, which is explained by strong excitation of density waves by shearing vortices at high Mach numbers. Note, however, 
that none of these authors considered the influence of bulk viscosity either setting it to zero, or adopting the Stokes hypothesis.

This paper is organised as follows. We start describing the 2D model for local perturbations of quasi-Keplerian shear flow in \S \ref{model_sect}. 
Next, in \S \ref{subsect4} we introduce the norm of perturbations and the definition of maximum optimal growth, $G_{max}$, to be determined further. 
We describe an iterative loop employed to obtain $G_{max}$. It consists of basic and adjoint dynamical equations, see equations (\ref{SFH_sys_4}-\ref{SFH_sys_6}) and (\ref{adj_SFH_sys_4}-\ref{adj_SFH_sys_6}), respectively.
Next, we proceed with a numerical study of the optimal growth. First, we analyse the suppression of $G_{max}$ by bulk viscosity 
and find that shearing vortices are almost unaffected by the value of bulk viscosity, see \S \ref{bulk_visc}. 
At the same time, the growth factors of large-scale vortices appear to be substantially larger than those of small-scale vortices. 
Secondly, in \S \ref{super_Kep} we present the optimal growth in super-Keplerian shear demonstrating that $G_{max}$ strongly depends on the shear rate, increasing as one proceeds to the constant specific angular momentum rotation. Moreover, 
this numerical result, see Fig. \ref{fig_5}, is in a good agreement with analytical estimations, relegated to Appendix \ref{prelim}.
In \S \ref{low_limit} we discuss possible consequences of our results speculating that the transition Reynolds number, as well as the dimensionless azimuthal turbulent stress, may strongly depend on the shear rate. Additionally, in \S \ref{global} we consider the optimal growth of large-scale shearing vortices on a global spatial scale, when the azimuthal wavelength of perturbations becomes comparable to a radial disc scale. It is shown that a global initial vortex is a radially localised leading spiral having a certain radial position and a degree of winding both altering with the optimisation timespan. Employing the general condition (\ref{r_sp}), we show that these properties of the leading spiral can be explained well in terms of the local approach even for perturbations with the azimuthal wavenumbers of order of unity. However, the value of globally determined $G_{max}$ substantially depends on particular radial profiles of surface density and viscosity in a flow with alternating shear rate, e.g., in a relativistic disc. 
We give a summary of our study and briefly discuss further developments in Conclusions.

\section{Model for dynamics of perturbations}
\label{model_sect}

In order to study the transient dynamics of perturbations taking into account a finite disc scale-height, we employ the most basic model of a perturbed flow.
As we noted in the introductory section, the most rapidly growing vortices in quasi-Keplerian flows have columnar structure, and the swing amplification of spirals occurs 
via the dynamics in the disc plane. That is why we consider the perturbations which are independent of the disc vertical direction. 
Next, we leave out the details of energy balance between viscous heat and radiative (or/and thermal conductivity) cooling
of fluid elements adopting a polytropic equation of state with polytropic index $n$ for both background and perturbed flows. Note that polytropic perturbations with velocity field independent of vertical coordinate
preserve hydrostatic equilibrium, i.e. they have planar velocity field. Thus, the 3D problem with such a restriction is reduced to 2D problem after one integrates equations along the vertical coordinate.  
However, one should bear in mind that vertical and planar perturbed motions can be fully separated from each other in the particular cases of vertically isothermal or vertically homogeneous flows only.
Otherwise, even the initially columnar perturbation may acquire vertical dependence and, consequently, vertical velocity perturbation during its evolution. This issue must be addressed in our future study of 
optimal transient growth of shearing vortices in vertically inhomogeneous baratropic quasi-Keplerian flow, from which the restriction of perturbation columnar structure is going to be removed, see also the discussion
in \S \ref{future}.

Substituting the polytropic relation between Eulerian perturbations of pressure and density, $p_1$ and $\rho_1$, as well as their
background counterparts, $p$ and $\rho$, into the continuity equation, we change to a new variable $W = p_1/\rho$ in the dynamical equations for linear perturbations. 
These equations are written in the large shearing box approximation (see A.3 in \citet{umurhan-regev-2004}) and integrated across the disc thickness. 
We finally get:

\begin{equation}
\label{sys1}
\left ( \frac{\partial}{\partial t} - q\Omega_0 x\frac{\partial}{\partial y} \right ) u_x - 2\Omega_0 u_y =
-\frac{\partial W}{\partial x} + f_x + g_x,
\end{equation}
\begin{equation}
\label{sys2}
\left ( \frac{\partial}{\partial t} - q\Omega_0 x\frac{\partial}{\partial y} \right ) u_y + 
(2 - q)\Omega_0 u_x =
-\frac{\partial W}{\partial y} + f_y + g_y,
\end{equation}
\begin{equation}
\label{sys3}
\left ( \frac{\partial}{\partial t} - q\Omega_0 x\frac{\partial}{\partial y} \right ) W + 
a_*^2 \left ( \frac{\partial u_x}{\partial x} + \frac{\partial u_y}{\partial y} \right ) = 0,
\end{equation}
where $u_x,u_y$ are the Eulerian perturbations of the velocity components excited in a small patch of the disc.
Variables $x$ and $y$ are Cartesian coordinates, which locally correspond to radial and azimuthal directions in the disc, whereas
$\Omega_0$ is angular velocity of the fluid rotation at $x=0$ corresponding to some radial distance $r_0\gg x$. Also, $q$ is a constant shear rate that defines 
the background azimuthal velocity as $v_y=-q\Omega_0 x$. All other background quantities, such as 
$\Sigma\equiv \int \rho dz$ and $a_*$, are assumed to be constant. Note that $a_*^2=n a_{eq}^2/(n+1/2)$ with $a_{eq}$ being sound speed
in the equatorial plane of disc. The inviscid part of (\ref{sys1}-\ref{sys3}) coincides with the set (28-30) given in \citetalias{zhuravlev-razdoburdin-2014}.
At the same time, the new ingredients are viscous terms, which we take as follows:
\begin{equation}
\label{sys_f}
f_{x,y} = \nu \left ( \frac{\partial^2}{\partial x^2} + \frac{\partial^2}{\partial y^2} \right ) u_{x,y}, 
\end{equation}
\begin{equation}
\label{sys_g}
g_{x,y} = (\nu_b + \nu/3) \frac{\partial}{\partial x,\partial y} \left ( \frac{\partial u_x}{\partial x} + \frac{\partial u_y}{\partial y} \right ),
\end{equation}
where $\nu$ and $\nu_b$ denote kinematic viscosity coefficients corresponding, respectively, to the shear and bulk viscosities of the background flow, 
$\eta$ and $\zeta$, integrated across the disc thickness and subsequently normalised by $\Sigma$. 

In order to obtain $f_{x,y}$ and $g_{x,y}$ in the form (\ref{sys_f}-\ref{sys_g}), we first collect perturbed parts of 
all dissipative terms entering the Navier-Stockes equations, see for example Appendix B in \citet{kato-2008}. 
Note that those terms emerge not only due to viscous forces but also as a consequence of radial advection in the background flow.
However, in the case of local perturbations with characteristic length $\lambda\ll r_0$, only the dissipative terms containing the highest order 
spatial derivatives of perturbations should be retained. This suggests that all contributions containing $\nu\rho_1/\rho$ or $\eta_1/\rho$, 
where $\eta_1$ is the Eulerian perturbation of dynamical viscosity, or proportional to the radial background velocity $v_x \sim \nu / r_0$ should be omitted.
Indeed, for example the advection term
$$
v_x \frac{\partial u_x}{\partial x} \sim \nu \epsilon_u (r_0 \lambda)^{-1},
$$
whereas one of the terms from (\ref{sys_f})
$$
\nu \frac{\partial^2 u_x}{\partial x^2} \sim \nu \epsilon_u (\lambda)^{-2},
$$
which is $r_0/\lambda$ times larger ($\epsilon_u$ is an amplitude of $u_x$).
In this way, we obtain expressions (\ref{sys_f}-\ref{sys_g}).

\subsection{Dimensionless equations for single SFH}
\label{subsect2}

Shear flow with $a_* < \infty$ has a finite scale-height $H = a_* / \Omega_0$ what follows from the condition of vertical hydrostatic equilibrium in unperturbed disc with a polytropic equation of state. 
Following \citetalias{zhuravlev-razdoburdin-2014}, let us use the natural $H$ in our choice of dimensionless comoving Cartesian coordinates:

\begin{equation}
\label{compr_coords}
x^\prime =  x/H,\, y^\prime = (y+q\Omega_0 xt)/H,\, t^\prime=\Omega_0 t.
\end{equation}

The coordinates (\ref{compr_coords}) lead us to a spatially homogeneous set of equations which in {\it the case of boundless shear} have partial solutions in the form
of SFH,
$
f = \hat f (k_x,k_y,t^\prime) \exp ({\rm i} k_x x^\prime + {\rm i} k_y y^\prime),
$
where $f$ is any of the unknown quantities, $\hat f$ is its Fourier amplitude and 
$(k_x,k_y)$\footnote{it is assumed everywhere below that $k_y>0$.} are the dimensionless wavenumbers along $x$ and $y$ axes expressed in units of $H^{-1}$.
Omitting the primes from now on and introducing the shearing radial wavenumber $\tilde k_x \equiv k_x + q k_y t$ as well as
the full wavenumber squared $k^2 \equiv \tilde k_x^2 +k_y^2$,  we arrive at the following dimensionless equations:

\begin{multline}
\label{SFH_sys_4}
\frac{d \hat u_x}{d t} = 2\hat u_y - {\rm i}\tilde k_x \hat W - R^{-1} k^2 \hat u_x - \\ (R^{-1}/3 + R_b^{-1}) \tilde k_x (\tilde k_x \hat u_x + k_y \hat u_y),
\end{multline}

\begin{multline}
\label{SFH_sys_5}
\frac{d \hat u_y}{d t} = -(2-q)\hat u_x - {\rm i} k_y \hat W - R^{-1} k^2 \hat u_y - \\ (R^{-1}/3 + R_b^{-1}) k_y (\tilde k_x \hat u_x + k_y \hat u_y),
\end{multline}

\begin{equation}
\label{SFH_sys_6}
\frac{d \hat W}{d t} = - {\rm i}\, ( \, \tilde k_x \hat u_x + k_y \hat u_y \,).
\end{equation}
In equations (\ref{SFH_sys_4}-\ref{SFH_sys_6}) it is assumed that $\hat u_x,\hat u_y$ are expressed in units of $a_*$ and $\hat W$ is expressed in units of $a_*^2$.
Also, there are two different Reynolds numbers in (\ref{SFH_sys_4}-\ref{SFH_sys_6}), $R$ and $R_b$, which parametrise dynamical and bulk viscosities, respectively. 
Explicitly,
\begin{equation}
\label{reynolds}
R\equiv a_*^2/(\nu\Omega_0), \quad R_b \equiv a_*^2 / (\nu_b \Omega_0).
\end{equation}

In Appendix \ref{subsect1} we describe the incompressible limit of the model presented above. Note that in this limit the wavenumbers are expressed in units of inversed auxiliary length, rather than in units of $H^{-1}$. 
Also, the dynamical problem becomes of the first order, i.e. for particular values of $k_x, k_y$, $q$ and $R$, one uniquely determines a normalised solution
and the corresponding transient growth factor, see equations (\ref{sol_incompr}) and (\ref{g_incompr}).
On the contrary, the problem (\ref{SFH_sys_4}-\ref{SFH_sys_6}) has the third order, that is why, one first has to find a particular solution of (\ref{SFH_sys_4}-\ref{SFH_sys_6}) 
corresponding to the highest transient growth factor for a specified time interval. This solution is referred to as the optimal one.
To find this optimal solution, in \S \ref{subsect4} we are going to supply the set of equations (\ref{SFH_sys_4}-\ref{SFH_sys_6}) with the corresponding set of adjoint equations.
After that we will be able to implement the iterative loop elucidated in \S 2 of \citetalias{zhuravlev-razdoburdin-2014}.

\section{Transient growth factor}
\label{subsect4}

In this section we describe how the transient growth of perturbations is quantified in our study. 
As soon as finite compressibility is taken into account, the size of perturbations is commonly measured by surface density of their acoustic energy, $E_{ac}$, 
resulting in the following expression for dimensionless quantities constituting the SFH:
\begin{equation}
\label{compr_norm}
E_{ac} = \frac{1}{2} (|\hat u_x|^2 + |\hat u_y|^2 + |\hat W|^2).
\end{equation}
The growth factor of an arbitrary SFH is then given by
\begin{equation}
\label{g_compr}
g(t) \equiv \frac{E_{ac}(t)}{E_{ac}(0)}.
\end{equation}

Equation (\ref{compr_norm}) implies that in the functional space of perturbations with ${\bf q}$ and $\tilde {\bf q}$ being its members the inner product $({\bf q}, \tilde {\bf q})$ reads:
\begin{equation}
\label{inner_prod}
({\bf q}, \tilde {\bf q}) = \frac{1}{2} (\hat u_x \hat{\tilde u}_x^* + \hat u_y \hat{\tilde u}_y^* + \hat W \hat{\tilde W}^*),
\end{equation}
where asterisk stands for complex conjugation and ${\bf q} \equiv \{\hat u_x,\hat u_y,\hat W\}$, $\tilde {\bf q} \equiv \{\hat{\tilde u}_x,\hat{\tilde u}_y,\hat{\tilde W}\}$.

In order to find perturbation yielding the highest growth factor, we use the iteration procedure, see \citetalias{zhuravlev-razdoburdin-2014} for details.
Each step of iteration procedure consists of the advance of perturbation state vector forward in time employing the dynamical equations
\begin{equation}
\label{sys}
\frac{\partial {\bf q}}{\partial t} = {\bf A}\, {\bf q},
\end{equation}
and the subsequent integration of adjoint equations
\begin{equation}
\label{adj_eqs}
\frac{\partial \tilde {\bf q}}{\partial t} = - {\bf A}^\dag \, \tilde {\bf q}
\end{equation}
backwards in time.
Here, operators $\mathbf{A}$ and $\mathbf{A}^{\dag}$ are related by the Lagrange rule:
\begin{equation}
\label{Lagr_rule}
(\tilde {\bf q},{\bf A}{\bf q}) = ({\bf A}^\dag \tilde {\bf q}, {\bf q}).
\end{equation}
The explicit form of ${\bf A}$ and ${\bf A}^\dag$ is defined by the particular model for perturbations and particular choice of their norm.
In the case when ${\bf A}$ is represented by equations (\ref{SFH_sys_4}-\ref{SFH_sys_6}) and the inner product is given by (\ref{inner_prod}), 
operator $\mathbf{A}^{\dag}$ corresponds to the following set of adjoint equations:

\begin{multline}
\label{adj_SFH_sys_4}
\frac{d \hat {\tilde u}_x}{d t} = (2-q)\hat {\tilde u}_y - {\rm i}\tilde k_x \hat{\tilde W} + R^{-1} k^2 \hat {\tilde u}_x + \\ (R^{-1}/3 + R_b^{-1}) \tilde k_x (\tilde k_x \hat {\tilde u}_x + k_y \hat {\tilde u}_y),
\end{multline}

\begin{multline}
\label{adj_SFH_sys_5}
\frac{d \hat {\tilde u}_y}{d t} = -2\hat {\tilde u}_x - {\rm i} k_y \hat {\tilde W} + R^{-1} k^2 \hat {\tilde u}_y + \\ (R^{-1}/3 + R_b^{-1}) k_y (\tilde k_x \hat {\tilde u}_x + k_y \hat {\tilde u}_y),
\end{multline}

\begin{equation}
\label{adj_SFH_sys_6}
\frac{d \hat {\tilde W}}{d t} = - {\rm i}\, ( \, \tilde k_x \hat {\tilde u}_x + k_y \hat {\tilde u}_y \,).
\end{equation}
Note that using equations (\ref{direct1}-\ref{direct3}) and the rule (\ref{Lagr_rule}) we derive another set of adjoint equations to consider the dynamics of global perturbations with $\lambda\sim r_0$, see \S \ref{adj_eqs}.

The set (\ref{adj_SFH_sys_4}-\ref{adj_SFH_sys_6}) looks rather similar to the set (\ref{SFH_sys_4}-\ref{SFH_sys_6}) except that the coefficients standing in front of $\hat u_{y,x}$ in the inviscid terms
of equations (\ref{SFH_sys_4}) and (\ref{SFH_sys_5}), respectively, are flipped over, and the viscous terms acquire an opposite signs in the adjoint equations. 
The latter implies that (\ref{adj_SFH_sys_4}-\ref{adj_SFH_sys_5}) describe viscous diffusion with a negative diffusion coefficient. However, 
since the adjoint equations are advanced backwards in time during the implementation of iterative loop, one does not face any sort of instability caused by a non-zero negative viscosity. 

In this way, starting from an arbitrary perturbation state vector ${\bf q}$ at $t=0$, we use the following procedure:
\begin{itemize}
\item[1)]
integrate equations (\ref{SFH_sys_4}-\ref{SFH_sys_6}) forward in time up to the instant $t>0$;
\item[2)]
substitute the resulting ${\bf q}(t)$ into equations (\ref{adj_SFH_sys_4}-\ref{adj_SFH_sys_6}) and integrate them backwards in time up to the instant $t=0$;
\item[3)]
renormalise the upgraded ${\bf q}(0)$ to the unit norm;
\item[4)]
repeat the previous three steps until ${\bf q}(0)$ converges with a required accuracy.
\end{itemize}

As it is discussed in detail by \citetalias{zhuravlev-razdoburdin-2014}, the final ${\bf q}(0)$ represents the optimal SFH for particular values of $k_x$, $k_y$, $R$, $R_b$ and $q$ 
which exhibits the highest possible transient growth factor 
\begin{equation}
\label{g_opt}
g_{opt}(t) \equiv \max_{\forall {\rm SFH}} g(t) 
\end{equation}
up to the time $t$ among all other SFH's.
We call $g_{opt}$ the optimal growth hereafter.

Provided that $k_y$, $R$, $R_b$ and $q$ are fixed, $g_{opt}$ has a global maximum over all $k_x$ and $t$ which we denote as
\begin{equation}
\label{G_max_compr}
G_{max} \equiv \max_{\forall k_x,\forall t}\, g_{opt},
\end{equation}
since this quantity is analogous to the maximum optimal growth obtained for 3D incompressible perturbations by \citetalias{mukhopadhyay-2005}.
Note that in Appendix \ref{subsect3} we define an incompressible counterpart of (\ref{G_max_compr}).

Additionally, we are going to look at an intermediate quantity
\begin{equation}
\label{G_compr}
G(t)\equiv \max_{\forall k_x} g_{opt}
\end{equation}
in the case of compressible perturbations.

In what follows, $G_{max}$ is determined using common numerical routines for search of extremum.
In the rest of the paper we are going to present a parametric study of $G_{max}$ depending on $k_y$, $R$ and $q$\footnote{We also show the growth factors of corresponding optimals at the instant of swing in 
order to extinguish the exited density waves, see \S\ref{super_Kep}.}.
Such a kind of nonmodal problem with the full account of viscous terms in dynamical equations is solved for the first time.
Our goal is to determine accurate constraints on $G_{max}$, covering {\it vortical} perturbations with $\lambda_y\gtrsim H$ along with the case $\lambda_y \ll H$ considered previously in the literature.
It is also important to check the strong dependence of $G_{max}$ on $q$ estimated analytically by \citetalias{zhuravlev-razdoburdin-2014} (see also the review by \citet{razdoburdin-zhuravlev-2015}, \citetalias{razdoburdin-zhuravlev-2015} hereafter), 
which is elucidated in Appendix \ref{prelim} of this work, see equation (\ref{anal_G_max_compr}).

\section{Results}

\subsection{Bulk viscosity: suppression of shearing density waves}
\label{bulk_visc}

\citetalias{zhuravlev-razdoburdin-2014} showed that, as far as $k_x<0$, the optimal SFH corresponds to a pure vortex, rather then to a shearing density wave or a mixture of both, see also \citet{bodo-2005}. This vortex is a leading 
spiral contracted and amplified by the background shear. In other words, as far as $k_x<0$, the optimisation procedure for determining $g_{opt}$ is equivalent to looking for a pure vortex. 
At the time $t_s$ called the instant of swing the leading spiral turns into a trailing one starting its decay. By definition, $\tilde k_x(t_s)=0$. 
Additionally, at $t=t_s$ the spiral excites a shearing density wave with an initial amplitude defined by $k_y$. 
This density wave starts growing because its frequency increases due to the trailing spiral being wound up by the shear. 
The strongest density waves are generated by vortices with $k_y\sim 1$, see \citet{heinemann-papaloizou-2009a}. On the one hand, 
as soon as $k_y\sim 1$, they significantly enhance the optimal growth $G(t)$ as defined by equation (\ref{G_compr}), but on the other hand, it is not clear what role they play
in production and maintenance of turbulence. Additionally, as density waves produce highly divergent velocity field, they must be strongly affected by value of bulk viscosity which is poorly known in astrophysical flows. 
For these reasons, we are going to focus on the dynamics of vortices only. However, in order to be confident that a reliable quantitative criterion will be chosen to conclude about 
the transient amplification of vortices irrespective to excitation of density waves, one should first examine the optimal solutions to general problem represented by equations (\ref{SFH_sys_4}-\ref{SFH_sys_6}).

\begin{figure}
\begin{center}
\includegraphics[width=9cm,angle=0]{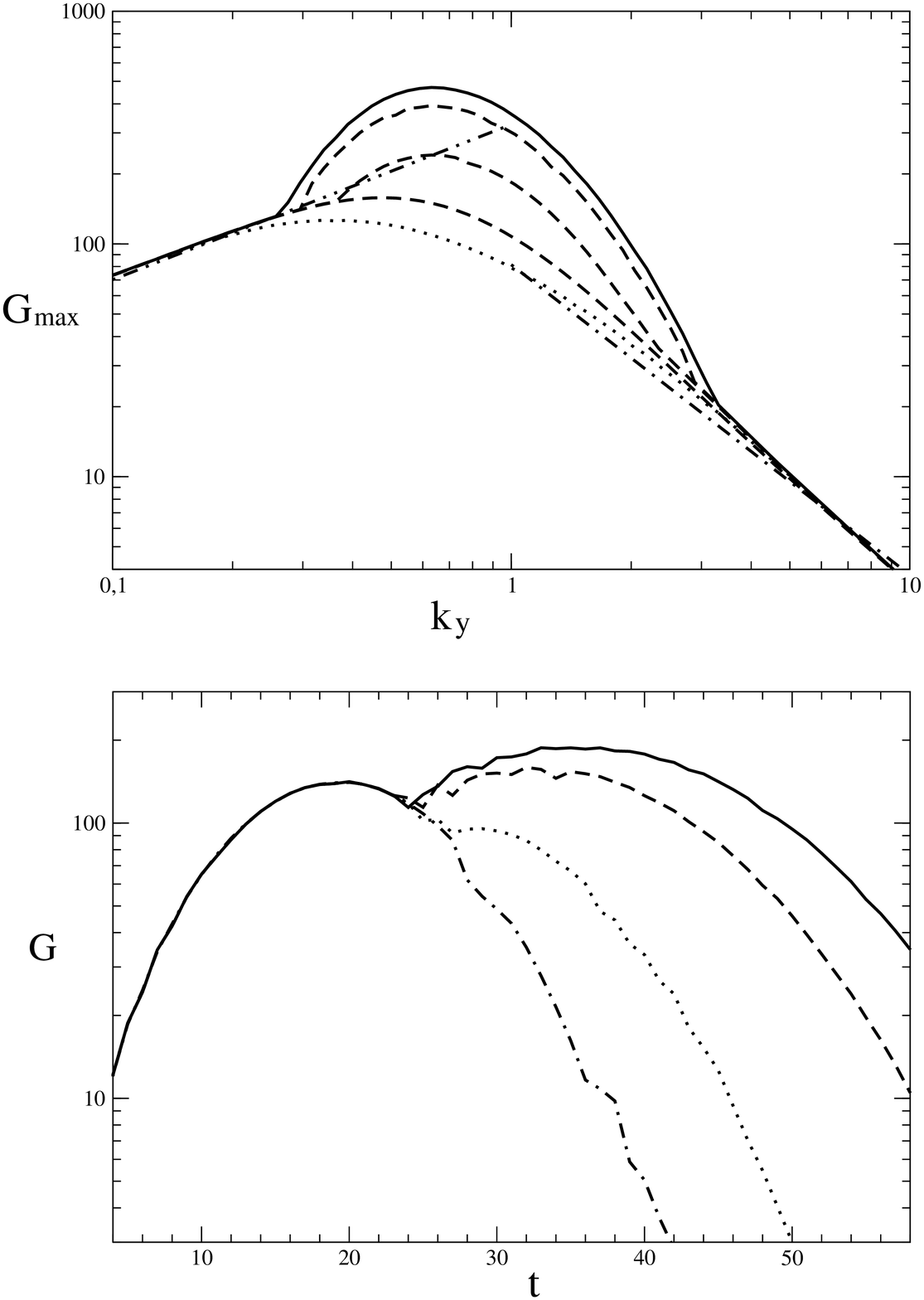}
\end{center}
\caption{Top panel: curves of $G_{max}$ as defined by equation (\ref{G_max_compr}) obtained via iterative loop for $R_{05}=2000$, $q=3/2$ and various $R_b$.
For solid curve $R_b=\infty$, for dashed curves top down $R_b=R, 0.1R, 0.01R$. 
The dotted curve shows $g(t_s)$ vs. $k_y$ for optimal SFH corresponding to $G_{max}$ represented by solid curve.
Dot-dashed and dot-dot-dashed curves are produced by equation (\ref{anal_G_max}) and equation (\ref{anal_G_max_compr}), respectively.
Bottom panel: curves of $G(t)$ as defined by equation (\ref{G_compr}) obtained via iterative loop for $R_{05}=2000$, $q=3/2$ and $k_y=0.3$.
Solid, dashed, dotted and dot-dashed curves represent $R_b=\infty, R, 0.1R, 0.01R$, respectively. }
\label{fig_23}
\end{figure}

In this regard, we present the result of optimisation procedure for the case of Keplerian shear in Fig. \ref{fig_23}.
For the sake of consistency with \citetalias{mukhopadhyay-2005} and with analytical estimations shown in Appendix \ref{prelim}, here we take $R_{05}=2000$. Note that $R_{05}$ is related to $R$ by equation (\ref{R_05}).
On the top panel in Fig. \ref{fig_23} one can see the highest possible enhancement of linear perturbations in the Keplerian shear which is plotted vs. $k_y$. 
First, we start with the case of zero bulk viscosity demonstrated by the solid curve and find the following:
\begin{itemize}
\item[1)]
despite the fact we consider boundless shear, it does have a maximum at $k_y\sim 1$ and wings declining for perturbations with both small and large azimuthal length scale in comparison with the disc thickness;
\item[2)]
these wings are matched well by the analytical expressions (\ref{anal_G_max}) and (\ref{anal_G_max_compr}) derived in Appendix \ref{prelim} employing the so called balanced solutions, see e.g. (35-37) by \citetalias{razdoburdin-zhuravlev-2015};
\item[3)] 
perturbations with $k_y < 1$ yield a considerably larger transient growth than previously analysed small-scale perturbations described in the incompressible limit, see Appendix \ref{incompr};
\item[4)]
in the range $0.3<k_y<3$ one finds a bump which apparently corresponds to the effect of density waves generation by vortices and is referred to as the 'sonic bump' hereafter.
\end{itemize}

The point (4) becomes more clear as we look at the bottom panel in Fig. \ref{fig_23}. In order to plot it, we fix $k_y=0.3$ which approximately corresponds to the left edge of the sonic bump for the case $R_b=\infty$
and obtain the profile of $G(t)$, see solid curve. In turns out that while approaching the area of $k_y\sim 1$, $G(t)$ acquires an additional local maximum which is located farther in time than $t_{max}$ given by equation (\ref{t_max}). 
So, $t_{max}\approx 19$ for the parameters taken to obtain the solid curve, which matches the location of its left maximum. If it were not for prominent excitation of density waves,
it would be a unique extremum of $G(t)$, see Fig. 9 by \citetalias{mukhopadhyay-2005}. Let us denote it by $G_{max}^\prime$. However, as it is shown by \citetalias{zhuravlev-razdoburdin-2014} in the inviscid model, see their Fig. 2, 
the vortices with $k_y\sim 1$ that have already swung well in advance of some fixed time interval $t_0$
produce an additional bump in the profile of $g_{opt}(k_x, t=t_0)$ due to their ability to generate sufficiently strong density waves. This additional bump is suppressed by non-zero viscosity
at a considerably later time than the local peak located at $k_x=-q k_y t_0$ and produced by vortices swinging right at $t=t_0$. As a consequence, $G(t)$ acquires the second local maximum, which we denote by $G_{max}^{\prime\prime}$. 
Further, as we approach $k_y\approx 1$ from either small $k_y$ or large $k_y$, we enter the domain where $G_{max}^{\prime\prime}> G_{max}^\prime$, or even $G_{max}^{\prime\prime}$ 'swallows up' $G_{max}^\prime$. 
This domain corresponds to the sonic bump in the solid curve on the top panel in Fig. \ref{fig_23}, where $G_{max} = G_{max}^{\prime\prime}$.

\begin{figure}
\begin{center}
\includegraphics[width=9cm,angle=0]{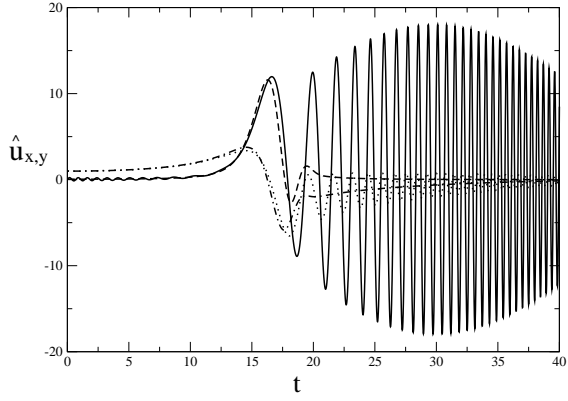}
\end{center}
\caption{
Solid and dashed curves are $\hat u_x(t)$ for optimal SFH corresponding to $G_{max}(k_y=0.4)$ obtained for $R_{05}=2000$, $q=3/2$ with $R_b=\infty$ and $R_b=0.001R$, respectively.
Dotted and dot-dashed curves are $\hat u_y(t)$ for the same two SFHs.
} \label{fig_4}
\end{figure}

Now, let us consider the non-zero bulk viscosity. On both panels in Fig. \ref{fig_23} another three curves of $G_{max}$ and three curves of $G(t)$ show the transient growth behaviour as we proceed from $R_b=\infty$ up to $R_b=0.01 R$.
Apparently, such a huge change in the value of bulk viscosity wipes out the density waves and the corresponding sonic bump but {\it does not affect} the growth of vortices. 
In order to see this better, we add the dotted curve on the top panel in Fig. \ref{fig_23}, which represents 
$$g_s\equiv g(t_s)$$ 
of vortical SFH that yields $G_{max}(k_y)$ for $R_b=\infty$. Clearly, in the limit $R_b \to 1$ the curve of $G_{max}$ just approaches the curve of $g_s$ from above. 
The same is also confirmed by the profiles of $\hat u_{x,y}(t)$ plotted in Fig. \ref{fig_4}. These velocity perturbations belong to the optimal vortices producing $G_{max}$ at $k_y=0.4$ in the cases of $R_b=\infty$ and $R_b=0.001R$. We find that the bulk viscosity enhancement leaves $\hat u_{x,y}$ almost unchanged for $t<t_s$, whereas their 
prominent oscillations disappear right after the instant of swing at $t_s\approx 16$. Also note that in the case of $R_b=\infty$ the oscillations of 
$\hat u_x$ start damping approximately at $t\approx 30$ which is the time corresponding to the value of optimal growth equal to $G_{max}^{\prime\prime}$. 
We do not present plots with $t_{max}(k_y)$, but it is checked that $t_{max}$ numerically determined for wings of $G_{max}(k_y)$ corresponding to $k_y\ll 1$ ($k_y\gg 1$) fits the analytical estimation (\ref{t_max}) very well.

The conclusion we have just made about the invariance of transient growth of vortical SFH with respect to the magnitude of bulk viscosity can be supported analytically in the case of tightly wound spirals $|\tilde k_x|\gg 1$, 
see Appendix \ref{bulk_app}. However, the latter assumption breaks down as soon as the optimal spiral unwinds and enters the swing interval, where $|\tilde k_x|\lesssim 1$, and one cannot even distinguish between vortices and density waves.
Note that the swing interval may be many times longer than $\sim\Omega_0^{-1}$ for large-scale vortices. That is why the numerical analysis presented above appears to be necessary to check the robustness of analytical conclusions.

Thus, in the next section we are going to use $g_s$ corresponding to perturbations exhibiting the maximum optimal growth, $G_{max}$, as a measure of the transient growth of shearing vortices.

\subsection{Transient growth in super-Keplerian shear of a finite scale-height}
\label{super_Kep}

\begin{figure}
\begin{center}
\includegraphics[width=8cm,angle=0]{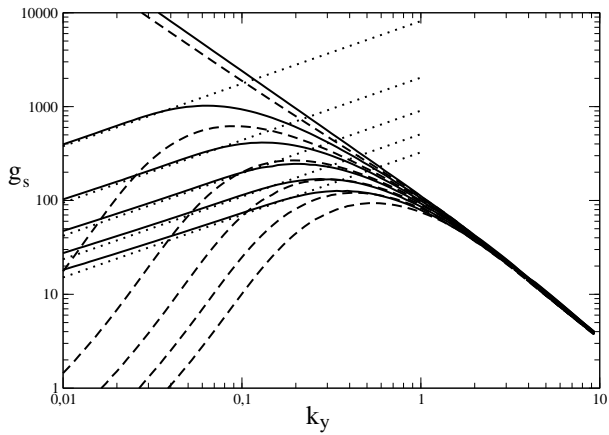}
\end{center}
\caption{Solid curves show $g_s$ vs. $k_y$ for optimal SFH that attains $G_{max}$ as defined by equation (\ref{G_max_compr}) for $R_{05}=2000$ and $R_b=\infty$.
Upwards $q=1.5, 1.6, 1.7, 1.8, 1.9, 2.0$. 
Dashed curves show $g_{kin}(t_s)$ vs. $k_y$ for the same SFHs. 
Upwards $q=1.5, 1.6, 1.7, 1.8, 1.9, 2.0$.
Dotted curves represent the analytical estimation (\ref{anal_G_max_compr}) for the same parameters and $q=1.5, 1.6, 1.7, 1.8, 1.9$ from bottom to top.
} \label{fig_5}
\end{figure}

As it is shown by \citet{butler-farrell-1992}, optimal vortices in an incompressible plane flow are streamwise rolls which transiently grow into streamwise streaks due to the 'lift-up' effect, see \citet{schmid-henningson-2001}. In the rotating constant specific angular momentum flow the optimal vortices are, oppositely, 
the streamwise streaks which generate streamwise rolls what has been referred to as the 'anti-lift-up' effect, see \citet{antkowiak-brancher-2007} and the discussion by \citet{rincon-2008}.
In both cases, the corresponding optimal growth $\propto R^2$ and attains several thousands for $R_{05}=2000$, see \citetalias{mukhopadhyay-2005}. In other words, this type of transient growth is produced by a nearly axisymmetric ($k_y\approx 0$) SFH 
of 3D perturbations and is limited only by viscous dissipation.
However, the growth of rolls in the rotating flow is dramatically suppressed by onset of epicyclic oscillations. The optimal growth drops steeply as one proceeds to the shear with a positive
radial gradient of the specific angular momentum. Indeed, according to the analytical result of \citetalias{mukhopadhyay-2005} $G_{max}\approx 2/(2-q)$, which gives, for example, 
$G_{max}\sim 20$ at $q=1.9$. At the same time, the swing amplification of columnar (independent on $z$) small-scale vortices $\lambda_y \ll H$ do not produce a substantial growth before they are damped by the viscosity, so that  
the corresponding maximum growth has value $G_{max} \sim 30$ at $q=1.9$ according to \citetalias{mukhopadhyay-2005}, which is only slightly larger than the value obtained for streamwise rolls.
Nevertheless, as we have seen in the previous Section, the swing amplification of 2D vortical perturbations with $\lambda_y \gtrsim H$ proves to be significantly larger than that of vortices with $\lambda \ll H$ in the Keplerian case $q=1.5$. 
Additionally, according to the analytical result (\ref{anal_G_max_compr}), the large-scale vortices should attain even larger growth factors as one proceeds to the super-Keplerian shear rate.
In other words, the swing amplification of large-scale vortices enhances as the stabilisation of fluid elements by Coriolis and tidal forces becomes weaker.
For this reason, we present results of numerical optimisation for $1.5<q<2.0$ involving the compressible dynamics in this Section, see Fig. \ref{fig_5}. 
There are solid curves that demonstrate gradual increase of growth factor as we approach $q=2$. 
As we see, at $q=1.9$ $g_s\sim 10^3$ which is two orders of magnitude larger than the corresponding incompressible result of \citetalias{mukhopadhyay-2005}. 
Also, these curves are matched well by their analytical counterparts, obtained using equation (\ref{anal_G_max_compr}), see the dotted lines in Fig. \ref{fig_5}. 
For a confidence, we numerically check that the dependence $g_s\propto R^{2/3}$ is valid throughout the whole range of $k_y$.
Thereby, in contrast to what we see in Fig. 4 of \citetalias{mukhopadhyay-2005}, the transient growth of perturbations in shear flow of a finite scale-height reduces much more gradually as one introduces a positive radial gradient of the specific angular momentum.

Besides, $g_s$ diverges as $k_y\to 0$ in the constant specific angular momentum flow, $q=2$. In other words, it obeys the same law as in the incompressible limit: $g_s\propto k_y^{-4/3}$. 
Using equation (57) from \citetalias{zhuravlev-razdoburdin-2014} at $\kappa=0$ we find that inviscid balanced solutions (35-37) of \citetalias{razdoburdin-zhuravlev-2015} yield $G_{max} = t_{max}^2$ in the limit $k_y\ll 1$ and $|k_x|\gg 1$, which also recovers the incompressible dependence on $k_y$ after we substitute $t_{max}$ (\ref{t_max}). However, the numerical curve for $q=2$ displayed in Fig. \ref{fig_5} gives a somewhat larger value in the limit $k_y\ll 1$, even though we have not included the factor of viscosity damping in the estimate of $G_{max}$. 
Presumably, this discrepancy is a consequence of prominent excitation of density wave up to $k_y\to 0$ which enhances the growth factor of optimal vortex comparing to the value coming from balanced solution at $t=t_s$. 
This claim is supported by the analytical result of \citet{heinemann-papaloizou-2009a} which states that the amplitude of generated density wave grows exponentially as the parameter $\epsilon\ll 1$ increases, see e.g. equation (34) of \citetalias{razdoburdin-zhuravlev-2015}. 
As long as $\kappa\sim 1$ and $k_y$ is not too close to unity, $\epsilon$ is small, the theory constructed by \citet{heinemann-papaloizou-2009a} is valid and the generated wave has a small amplitude at $t=t_s$.
That is why the numerically obtained $g_s$ is close to the estimate coming from the balanced solution for $k_y>1$ as well as for $k_y<1$.
However, as  $\kappa\to 0$, the region of weak wave excitation shifts to smaller $k_y$ (one can compare the pairs of solid and dotted curves for $q=1.5$ and $q=1.9$ in Fig. \ref{fig_5}).
For $q=2$ the whole region $0<k_y\lesssim 1$ suffers strong excitation of density waves, making the use of the balanced solution unjustified\footnote{This explanation should be taken with a caution that 
one cannot distinguish between vortices and waves in the swing interval and, in particular, at the instant of swing. Strictly, we can only say that $g_s$ becomes considerably larger than the estimate extracted from the balanced solution 
as soon as $\epsilon$ is not small. This remark is not unnecessary because, as we have already checked, $g_s$ is {\it not} affected by high bulk viscosity, $R_b\ll R$, even for $q=2$ in the region $\epsilon\sim 1$}. 
Though, still it reproduces the correct dependence on $k_y$.

\subsubsection{Kinetic growth factor}

\begin{figure}
\begin{center}
\includegraphics[width=9cm,angle=0]{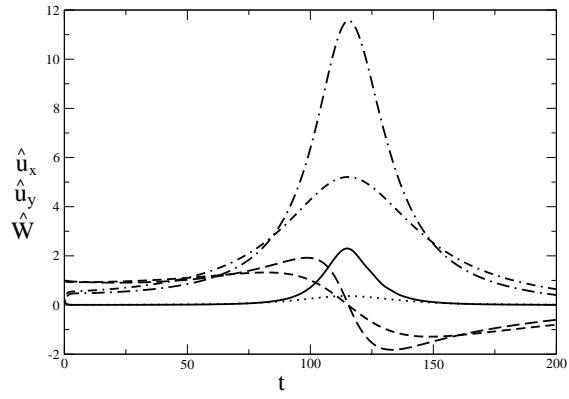}
\end{center}
\caption{The comparison of $\hat u_{x,y}$ and $\hat W$ for Keplerian and super-Keplerian shear.
Dotted, short-dashed and short-dashed-dotted curves show $\hat u_x$, $\hat u_y$ and $\hat W$ vs. $t$ for optimal SFH producing $G_{max}(k_y=0.02)$ as defined by equation (\ref{G_max_compr}) for 
$R_{05}=2000$, $R_b=0.001 R$ and $q=1.5$.
Solid, long-dashed and long-dashed-dotted curves show $\hat u_x$, $\hat u_y$ and $\hat W$ vs. $t$  for optimal SFH producing $G(t)$ at $k_y=0.02$, where $q=1.8$, $R_{05}=2000$, $R_b=0.001 R$ and $t$ equals to the swing time of the previous SFH. } \label{fig_6}
\end{figure}

Going back to the case $q<2$, the question arises for both Keplerian and super-Keplerian shear rates, to what extent either the velocity perturbation or the enthalpy perturbation provide the growth of $g_s$ as one proceeds to 
large azimuthal scales. To elucidate this, we additionally plot the kinetic part of the growth factor defined as
\begin{equation}
\label{g_kin}
g_{kin} = \frac{E_{kin}(t)}{E_{kin}(0)},
\end{equation}
where $E_{kin} = 1/2 (|\hat u_x|^2 + |\hat u_y|^2)$, see the dashed curves in Fig. \ref{fig_5}.
As we see, $g_{ks}\equiv g_{kin}(t_s)$ of the optimal large-scale vortices ceases much faster than $g_s$ as one shifts to $k_y\to 0$, i.e. a very large part of $g_s$ is contributed by the amplitude of $\hat W$. 
Nevertheless, it is interesting to note that the ratio $g_{ks}/g_s$ substantially increases as we fix $k_y$ and approach $q=2$. This can be seen also in Fig. \ref{fig_6}, where
we compare the profiles of $\hat u_x$, $\hat u_y$ and $\hat W$ in the Keplerian and super-Keplerian cases. The amplitude of $\hat u_x$ becomes larger by more than 6 times at the instant of swing, as we change from 
$q=1.5$ to $q=1.8$, whereas, the amplitude of $\hat W$ increases by a factor of two only. But still the kinetic energy is only a few percents of the acoustic energy of optimal SFH at the peak of its norm in the particular case
introduced in Fig. \ref{fig_6}\footnote{Additionally, we have checked that change to norm of perturbations with term $|\hat W|^2$ in equation (\ref{compr_norm}) multiplied by a constant $p\in(0,1]$ and corresponding change of 
the adjoint equations leads to the same optimal solutions being the leading spirals of the vortical SFH with negative $k_x$.}.

We are not the first who find the optimal transient perturbations with a significant or even prevailing deposit of enthalpy (alternatively, density or temperature) perturbation amplitude. 
A number of studies devoted to inspection of compressible transient perturbations in plane shear flows have found this previously. 
For example, \citet{hanifi-1996}, who showed that the lift-up mechanism survives in a moderately supersonic boundary layer, and optimal perturbations are independent of the streamwise direction similarly to the incompressible case, 
found also that these optimals contain large amplitude of temperature perturbation, see their Fig. 12. 
Next, \citet{farrell-ioannou-2000} solved a 2D optimisation problem in a hypersonic plane shear flow, which is more analogous to our model. Similarly to our results, they obtained that perturbation harmonics in the form of 
the leading spirals exhibit growth factors much larger than they would be in incompressible fluid because of the density waves excitation at the instant of swing. For the particular situation they considered 
the initial optimal perturbation mostly consisted of a pressure perturbation, see their Fig. 4. 
This general result is not surprising, at least in the case of 2D dynamics studied by \citet{farrell-ioannou-2000} and by us, since balanced solutions acquire the non-zero
velocity divergence and non-zero density perturbation with an account of compressibility, see equations (35-37) by \citetalias{razdoburdin-zhuravlev-2015}. It should be kept in mind, however, that the balanced solutions (as well as their accurate numerical counterparts) 
are indeed vortices since they are proportional to $I_\nu$ and are not affected by bulk viscosity, see \S {\ref{bulk_visc}}. 
Also note that 2D optimal growth in a hypersonic plane flow obtained by \citet{farrell-ioannou-2000} would be effectively suppressed by high bulk viscosity down to the incompressible value which is much smaller than the 3D optimal 
growth produced by the streamwise rolls. On the contrary, this is not the case in super-Keplerian rotating flow with $q\to 2$ since we find the enhancement of transient growth of large-scale vortices irrespectively of an excitation of
density waves, see Fig. \ref{fig_5}. Thus, for a super-Keplerian shear rate close enough to $q=2$ 2D transient growth produced by the swing amplification of shearing spirals exceeds 3D transient growth of streamwise streaks 
produced by the anti-lift-up effect. Note that it is correct to compare the growth factors of both types of perturbations since for $q=2$ the streamwise-independent solution in the form of the streaks, see equations (52-55) of \citetalias{mukhopadhyay-2005}, satisfies the set of equations (\ref{SFH_sys_4} - \ref{SFH_sys_6}) provided that the equation describing the evolution of the vertical velocity perturbation $\hat u_z$ is added. 
Also, it can be checked that such set of equations allows the existence of an axisymmetric vortical solutions for $q\lesssim 2$, which are akin of streamwise streaks with a non-modal growth strongly modified by epicyclic motions
and slightly modified by divergent motions up to $\lambda\sim H$.   

Finally, we argue that it is questionable which way the high amplitude density perturbations emerged from the transient growth of vortices with $\lambda_y \gtrsim H$ should affect the stage of non-linear interaction of different SFHs. 
Unlike density perturbations of the excited density waves, they hardly can lead to generation of shocks, because at the peak of their amplitudes, i.e. near the vortex swing, they have the least spatial gradient and, 
oppositely, as soon as their spatial gradient starts to grow through the spiral winding, their amplitude drops back to smaller values. Contrary to this, the amplitude of $W$ in density wave grows simultaneously with its spatial gradient, 
what most likely ends with a shock. It is tempting to suppose that high amplitudes of $W$ generated by large-scale vortices qualitatively change the picture of the non-linear dynamics 
through the term $\nabla \rho_1 {\bf u}$ entering the continuity equation for finite amplitude perturbations. However, to the best of our knowledge, it is not clear, whether this has any consequences with regards to 
the problem of positive non-linear feedback in the bypass transition to turbulence. To date, a significant progress has been made on wind tunnel experiments (e.g. \cite{schneider-2006}) 
and on direct numerical simulations of the transition in hypersonic boundary layers, see the review by \citet{zhong-wang-2012} and one of the recent particular examples by \citet{kudryavtsev-2015} . 
Further, theoretical studies has been carried out with emphasise on the transition scenario based on spectral instabilities of the hypersonic flow, see the review by \citet{fedorov-2011}. 
At the same time, we are not aware of any research analysing possible qualitative changes in the transition scenarios involving the bypass mechanism.

Nevertheless, whatever part of transient growth factor is contained in enthalpy perturbation, the kinetic part, $g_{ks}$, remains to be significant for optimal SFH with $k_y\lesssim 1$ even in the Keplerian case. 
Not less importantly, $g_{ks}$ gets larger as $q\to 2$ even more steeply than $g_s$ attaining the value $g_{kin}\sim 600$ for $q=1.9$ due to the substantial increase of $\hat u_x$ amplitude.

\subsection{Low limit of transition Reynolds number}
\label{low_limit}

\begin{figure}
\begin{center}
\includegraphics[width=8cm,angle=0]{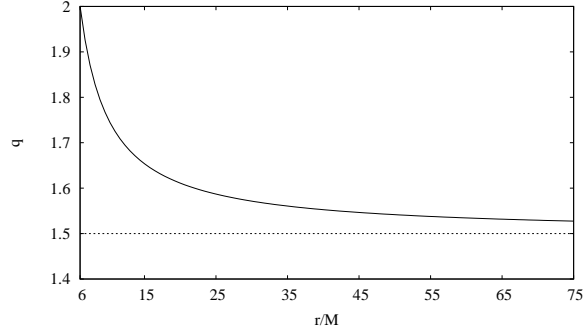}
\end{center}
\caption{The dependence of local shear rate $q$ on radial distance in the Paszynski-Wiita potential, see equation (\ref{q_r}). } \label{fig_7a}
\end{figure}

\begin{figure}
\begin{center}
\includegraphics[width=9cm,angle=0]{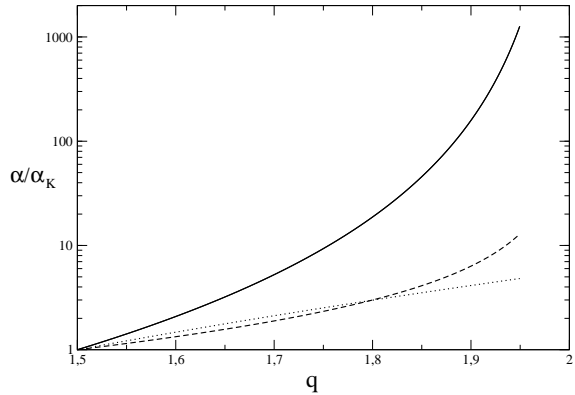}
\end{center}
\caption{The dependence of $\alpha$ on local shear rate $q$, see \S \ref{low_limit}. The solid curve represents equation (\ref{alpha}). The dashed and the dotted curves correspond 
to the results of \citet{abramowicz-1996-alpha} and \citet{penna-2013}, respectively.} \label{fig_7}
\end{figure}

\citet{meseguer-2002} found a strong correlation between $G_{max}$ for 3D incompressible perturbations and subcritical stability boundary in a viscous counter-rotating spectrally stable Taylor-Couette flow. 
Recently, \citet{maretzke-2014} suggested that the threshold for transition in any regime of Taylor-Couette rotation 
should be $G_{max}\sim 100$ basing on a variety of new experimental results and nonmodal growth calculations. Of course, it is not clear yet, whether this is true for the quasi-Keplerian regime, since to date, nobody has detected its turbulent state. Even more, it is not known, whether the 
transition threshold for $G_{max}$ exists in the case of vortices in compressible medium considered in this work. Nevertheless, let us consider a conjecture about the value of the transition Reynolds number resulting from the assumption that  
$G_{max}(R_{T}) \sim const$, where $R_{T}$ is the Reynolds number of the subcritical transition in quasi-Keplerian case $1.5 \leq q < 2$. 
If so, equation (\ref{anal_G_max_compr}) yields
\begin{equation}
\label{R_q}
R_{T} \sim 10^8 \, \frac{(2-q)^3}{q},
\end{equation}
where we set $R_{T}\sim 10^7$ for the Keplerian rotation which is stable at least up to the Reynolds number equal to a few millions, see \citet{edlund-ji-2014}.

According to equation (\ref{R_q}), the bypass transition to HD turbulence may occur at significantly lower $R$, for example, in the inner region of relativistic accretion disc.
Indeed, using the Paczynski-Wiita potential \citep{paczynsky-wiita-1980}
\begin{equation}
\label{Paczynski}
\Phi_{PW} = - \frac{M}{r-2M},
\end{equation}
where $M$ is the mass of central object, $r$ is the radial coordinate in the disc plane and it is implied that $G=c=1$, we obtain that
\begin{equation}
\label{q_r}
q \equiv \frac{r}{\Omega}\frac{d\Omega}{dr}= \frac{1}{2} + \left ( 1-\frac{2M}{r}\right )^{-1}.
\end{equation}
Equation (\ref{q_r}) is plotted in Fig. \ref{fig_7a} which shows that for $r<20M$, where relativistic disc emits most its energy, $q>1.6$ and $R_{T}\lesssim 4\cdot 10^6$.

Next, we note that kinematic viscosity is likely to increase with the radial distance in the standard thin disc model:  
\begin{equation}
\label{nu_estim}
\nu \sim v_p l_p \propto a_* / N \sim \Omega H^2 / \Sigma \propto r^{1/2+\beta},
\end{equation}
where $v_p$, $l_p$ and $N$ are, respectively, the mean velocity, the mean free path and the number density of gas particles, and we assume that disc has a constant aspect ratio $H/r$ and $\Sigma\propto r^{-\beta}$ with $\beta >0$\footnote{In equation (\ref{nu_estim}) it is assumed that cross section of collisions between the particles is 
independent of $a_*$ and $N$.}. 
Consequently, the situation is possible when relativistic disc is turbulent inside some radius $r=r_T$, whereas turbulence ceases in the outer regions of disc. 
Presumably, if this happens, the angular momentum flux is absent throughout the laminar flow, and the inner disc exhibits a non-stationary accretion. 
Whether this process may become recurrent, is unclear and requires an additional research.  

Furthermore, it is known that in a disc with super-Eddington accretion rate the radial pressure gradient becomes significant flattening the specific angular momentum distribution of the rotating matter, see \citet{abramowicz-1988}.
Thereby, actual $q(r)$ should become even steeper comparing to the curve shown in Fig. \ref{fig_7a}.  
Of course, the enhanced transient growth of large-scale vortices may drive turbulisation and associated angular momentum outflow in the Newtonian discs, 
where the super-Keplerian rotation may arise solely due to the radial pressure gradient. 
More generally, the aforementioned mechanism may be at work whenever the super-Keplerian rotation takes place on streamwise scales comparable to or exceeding the vertical scale-height in a flow.
Thus, in principle, a kind of a secondary bypass transition can occur inside the elongated non-linear vortices, which produce a sufficient excess of vorticity on a smooth background shear they are embedded in.
Those non-linear vortices have been found in a large number of simulations including mentioned in introductory section \citep{umurhan-regev-2004,lithwick-2009}.
If elongated in the streamwise direction, they are known to be almost linearly stable, see \citet{lesur-papaloizou-2009}, and are efficient in trapping the dust particles in protoplanetary dics, see e.g. \citet{kenyon-2010}.

Another consequence of the dependence of $G_{max}$ on $q$ takes place for the dimensionless azimuthal viscous stress $\alpha$ if it is caused by turbulent motions fed by transiently growing large-scale shearing vortices.
\citet{lesur-longaretti-2005} suggested that $\alpha \propto R_{T}^{-1}$, thus, $\alpha$ should change with $q$ like
\begin{equation}
\label{alpha}
\alpha/\alpha_K \sim \frac{q}{12 (2-q)^3},
\end{equation}
where $\alpha = \alpha_K$ is its Keplerian value. 
Local and global MHD simulations of relativistic accretion discs, see e.g. \citet{penna-2013} and \citet{teixeira-2014}, show a radially varying $\alpha$ decreasing to the periphery of disc,
though, this is what one finds in a supercritical turbulence fed by MRI. 
In Fig. \ref{fig_7}, we compare equation (\ref{alpha}) with the expressions given in equation (9) by \citet{abramowicz-2009} and in the abstract of \citet{penna-2013}.
Since $q$ increases as we approach the inner edge of disc, see equation (\ref{q_r}), 
we suppose that if the subcritical turbulence based on the bypass mechanism and fed by transiently growing quasi-2D optimal shearing vortices were resolved in simulations, one would obtain  
a steeper dependence of $\alpha$ on $r$ comparing to what has been obtained to date with regards to supercritical magnetic turbulence.
In the case of relativistic disc, such a substantial radial variation of $\alpha$ on $r$ would result in a significant drop in temperature 
of the inner parts of disc for a fixed accretion rate $\dot M$, thus, increasing the limiting value of $\dot M$ corresponding to the Eddington luminosity in the standard disc model. 
According to equation (\ref{alpha}), $\alpha$ diverges in the vicinity of the inner edge of standard relativistic disc. However, the finite aspect ratio of disc, $\delta = H/r_0$, puts a lower limit on the value of  $k_y\sim\delta$, what constrains the value of optimal growth as $q\to 2$, see Fig. \ref{fig_5}. Additionally, the non-zero pressure radial gradient close to the inner edge of standard disc
always results in the shear rate $q<2$.

\section{Extension to global spatial scale}
\label{global}

So far, we have been considering transient dynamics of large-scale shearing vortices in the local framework implying that $r_0 \gg\lambda_y\gtrsim h$.
However, as $k_y\sim\delta$, one has to take into account a global disc structure and the cylindrical geometry of the problem. 
At first, disc generally is radially non-uniform as its surface density and its microscopic viscosity vary on the scale $\sim r_0$.
Next, the transient amplification of large-scale vortices substantially increases along with the local shear rate, which in turn may be a function of the radial distance. 
For example, in relativistic disc model with the Paczynsky-Wiita gravitational potential $q(r)$ is given by equation (\ref{q_r}).
Additionally, even if $q$ is a constant throughout the disc, which is the case for the Newtonian potential, global perturbations may be affected by a non-zero 
radial gradient of the vorticity, which vanishes in the local spatial limit. 

We have found in this work that the above factors may considerably alter a value of $G_{max}$ for global shearing vortices or even 
make the flow unstable. In order to see this, we present a supplementary study of global optimal perturbations 
using a set of global disc models with two variants of surface density and microscopic viscosity radial distributions in both 
the Newtonian and the Paczynsky-Wiita gravitational potentials. At the same time, we finally come to overall conclusion that regardless of 
a particular radial structure of disc it is always able to generate prominent transient growth of global perturbations at huge Reynolds numbers 
realised in astrophysical conditions.

\subsection{Background models}
\label{background}

For the first variant of the background we would like to take profiles of surface density, disc semi-thickness and disc radial velocity 
which correspond to a stationary accretion with a radially constant $\dot M$ occurring due to the fluid microscopic viscosity.

\subsubsection{Stationary configuration}
\label{basic_background}
Geometrically thin disc in the Paczynsky-Wiita potential (\ref{Paczynski}) rotates with the angular frequency
\begin{equation}
\label{Omega2}
\Omega = \frac{M^{1/2}}{r^{1/2}\left(r-2M\right)}
\end{equation}
and is subject to radial oscillations with the following epicyclic frequency
\begin{equation}
\label{kappa2}
\kappa = \frac{M^{1/2}}{\left(r-2M\right )^{3/2}}\left(1 - \frac{6M}{r}\right)^{1/2}.
\end{equation}

Another quantity entering equations for perturbations is the sound speed at the equatorial plane of disc, $a_{eq}$.
Provided that $\rho(z=H)=0$ and disc is polytropic along the vertical direction, there is the following exact relation between
the disc thickness and the equatorial speed of sound:
$$
\label{a_eq}
a_{eq}=\frac{H\Omega}{(2n)^{1/2}}.
$$
Since in our 2D problem we are dealing with vertically integrated equations for perturbations, which contain $a_*^2=n a_{eq}^2/(n+1/2)$, see \citetalias{zhuravlev-razdoburdin-2014}, we obtain
\begin{equation}
\label{a_glob}
a_* = \frac{H\Omega}{(2n+1)^{1/2}}.
\end{equation}
Next, one has to specify the kinematic viscosity coefficients, $\nu$ and $\nu_b$, across the flow.
Similarly to what has been defined in \S \ref{model_sect}, those are the shear and the bulk viscosities, $\eta$ and $\zeta$, integrated over the disc thickness and subsequently divided by surface mass density, $\Sigma$. As we have mentioned above, see equation (\ref{nu_estim}), it is reasonable to assume that $\nu\propto H^2\Omega/\Sigma$.
Provided that the value of $\nu$ is known at some distance $r=r_1$, 
\begin{equation}
\label{nu_r}
\nu(r)=\nu_1 \frac{\Sigma_1}{H_1^2\Omega_1} \frac{H^2(r)\Omega(r)}{\Sigma(r)},
\end{equation}
where the variables marked by the subscript '1' stand for the corresponding quantities taken at $r_1$. 
Let us choose $r_1$ being the distance, where $\nu(r)$ attains the minimum.
A value of $\nu_1$ can be specified via the Reynolds number $R$ (cf. equation (\ref{reynolds})): 
\begin{equation}
\label{Re}
R = \frac{a_{*1}^2}{\Omega_1 \nu_1} = \frac{H_1^2 \Omega_1}{\nu_1 (2n+1)}.
\end{equation}

Combining (\ref{nu_r}) with (\ref{Re}) we obtain the final expression for the kinematic viscosity coefficient profile:
\begin{equation}
\label{nu_r_Re}
\nu = \frac{q \Sigma_1}{(2n+1)R_{05}}\frac{H^2(r)\Omega(r)}{\Sigma(r)},
\end{equation}
where it is taken into account that $R_{05}=q(r_i) R$, see \S \ref{subsect1}, and $r_i$ is the inner boundary of disc.

With regards to the second viscosity, we assume that the second (local) Reynolds number is constant throughout the disc, which implies that
\begin{equation}
\label{zeta}
\nu_b= \frac{a_{*}^2}{\Omega R_b}.
\end{equation}

At the next stage we are going to get an expressions for the background radial velocity, $v_r$, and the disc half-thickness $H$.
For this purpose we use the azimuthal projection of the Navier-Stokes equations integrated over $z$
\begin{equation}
\label{v_r1}
\Sigma v_r \frac{d}{dr} (\Omega r) + \Sigma v_r \Omega=\frac{1}{r^2}\frac{d}{d r}\left(\Sigma \nu r^3 \Omega^{\prime}\right)
\end{equation}
and mass conservation in the form
\begin{equation}
\label{v_r2}
v_r=-\frac{\dot M}{2\pi \Sigma r}
\end{equation}
with $\dot M$ constant throughout the disc.

Equation (\ref{v_r2}) allows us to integrate (\ref{v_r1}) assuming that $r\varphi$-component of the stress tensor vanishes at the inner boundary of disc $r=r_i$:
\begin{equation}
\label{v_r3_integrated}
\dot M \Omega r^2 D=-2\pi \Sigma \nu r^3 \Omega^{\prime}.
\end{equation}
In equation (\ref{v_r3_integrated}) we introduce $D=1-\left(\Omega_i r_i^2\right)/ \left(\Omega r^2\right)$.
Note that for the Newtonian potential $D=1-(r_i/r)^{1/2}$, whereas for the Paczynsky-Wiita potential 
$D=1-\left(r_i/r\right)^{3/2}\left(r-2M\right) / \left(r_i-2M\right)$.

Combining (\ref{v_r2}) with (\ref{v_r3_integrated}) we obtain the radial velocity
\begin{equation}
\label{v_r4}
v_r=\frac{\nu \Omega^{\prime}}{\Omega D}.
\end{equation}

Then, combining (\ref{nu_r}) with (\ref{v_r3_integrated}) allows us to obtain the radial profile of $H$:
\begin{equation}
\nonumber
H\propto\left(-\frac{D}{r\Omega^{\prime}}\right)^{1/2}
\end{equation}

In order to parametrise $H$, we use the disc aspect ratio $\delta\equiv (H/r)|_{r=r_2}$ with $r_2$ being the distance, where $H/r$ takes its maximum value.
Finally,
\begin{equation}
\label{H_r}
H=\delta r_2\left(\frac{r_2 \Omega_2^{\prime}}{r \Omega^{\prime}}\right)^{1/2}\left(\frac{D}{D_2}\right)^{1/2},
\end{equation}
where the variables marked by the subscript '2' denote the corresponding quantities taken at $r_2$. 

At last, in order to specify the radial profile of surface density for such a disc we proceed similar to \cite{shakura-sunyaev-1973} and
assume that viscous dissipation is locally balanced by the radiation flux from the disc surface:
\begin{equation}
\label{S-S_2.6}
Q=\frac{1}{2}\Sigma \nu r^2 \left(\Omega^{\prime}\right)^2.
\end{equation}
From the other hand, the diffusive radiation transfer in vertical direction yields an approximate relation between the surface and the equatorial temperatures,
$T_{eff}$ and $T_{eq}$:
$$
T_{eq}^4 / T_{eff}^4 \propto \tau = H / l_{ph},
$$
where $l_{ph}\propto \rho^{-1}$ is the mean free path of photons in the case when the electron scattering prevails the free-free absorption opacity.
Thus, assuming that the non-relativistic gas pressure prevails, one concludes that
\begin{equation}
\label{S-S_2.7}
T_{eq}^4 \propto a_{eq}^8 \propto Q \Sigma.
\end{equation}

Combining (\ref{nu_r_Re}), (\ref{H_r}), (\ref{S-S_2.6})  and (\ref{S-S_2.7}) we finally obtain
\begin{equation}
\label{sigma}
\Sigma=C_\Sigma\frac{D^3\Omega^7}{r^5 (-\Omega^{\prime})^5}
\end{equation}
where $C_{\Sigma}$ is a dimensional constant.
Since the problem for dynamics of perturbations is linear, an explicit form of $C_\Sigma$ is unimportant for us.

Equations (\ref{Omega2}), (\ref{kappa2}), (\ref{a_glob}), (\ref{nu_r_Re}), (\ref{zeta}), (\ref{v_r4}), (\ref{H_r}) and (\ref{sigma}) specify our stationary background configuration.
Everywhere below we denote it as 'P1' in the case of $\Omega$ and $\kappa$ given by equations (\ref{Omega2}) and (\ref{kappa2}), otherwise, i.e. in the limit $r\gg 1$
when $\Omega$ and $\kappa$ acquire their Newtonian values we denote the background as 'N1'.

\subsubsection{Homogeneous configuration}
\label{simple_background}
The models N1 and P1 describe the quasi-stationary stage of accretion which can be achieved by the inner region of an arbitrary rotating laminar flow with $\delta\ll 1$ 
evolving due to microscopic viscous relaxation only. However, someone would argue that this stage is quite unrealistic, since microscopic 
viscosity is too small in typical astrophysical situation leading to a very long relaxation time which most likely exceeds the 
turbulisation time of any initial rotating configuration by orders of magnitude.
For this reason we regard the models N1 and P1 just as the limiting case which we would like to counterpose to a simplified radially homogeneous 
configuration described below.

As before, the frequencies and the sound speed are given by equations (\ref{Omega2}), (\ref{kappa2}), (\ref{a_glob}).
However, now the surface density is assumed to be a constant
\begin{equation}
\label{sigma_simple}
\Sigma=const.
\end{equation}

Accordingly, we assume that kinematic viscosity coefficient is given by (cf. equation (\ref{nu_r_Re}))
\begin{equation}
\label{nu_simple}
\nu= \frac{a_{*}^2}{\Omega R}.
\end{equation}
The bulk viscosity coefficient is given by equation (\ref{zeta}) as before.
The half-thickness profile is chosen to be
\begin{equation}
\label{H_simple}
H=\delta r
\end{equation}
Note that in a disc of such simple shape the locally defined dimensionless azimuthal wavenumber at every distance corresponds to 
a global harmonic of perturbations with one and the same azimuthal wavenumber, see equation (\ref{local_k_y}) below. 
At last, the radial velocity is set to zero
\begin{equation}
\label{v_r_simple}
v_r=0.
\end{equation}

Thereby, equations (\ref{Omega2}), (\ref{kappa2}), (\ref{a_glob}), (\ref{nu_simple}), (\ref{zeta}), (\ref{sigma_simple}), (\ref{H_simple}) and (\ref{v_r_simple}) specify our 
homogeneous configuration. Similarly, everywhere below we denote this configuration as 'P2' in the case of $\Omega$ and $\kappa$ corresponding to the Paczynsky-Wiita potential, whereas for the Newtonian case the model is denoted as 'N2'.

\subsection{Equations for perturbations}

In order to derive the set of equations for global perturbations we linearise the Navier -- Stokes equations written in the cylindrical coordinates $\{r,\varphi,z\}$ 
(see e.g. Appendix B of \cite{kato-2008}).
Following \S \ref{model_sect} we do not consider the energy balance in viscous fluid adopting the equation of state in the perturbed flow to have the same polytropic index as in the background equation of state. 
Similarly to what has been done in the local case, we restrict ourselves considering the planar velocity perturbations independent on vertical coordinate.
Thus, perturbations are described by three variables: the Eulerian perturbations of velocity components, $\delta v_r$ and $\delta v_\varphi$, 
and the Eulerian perturbation of enthalpy $\delta h$.
Owing to the rotational symmetry of background, it is sufficient to consider complex amplitudes of the Fourier harmonic $\propto e^{{\rm i} m\varphi}$ 
which depend on $r$ and $t$ only.
Integrating equations over $z$ we obtain the non-viscous terms to be exactly the same as in section 3.1 of \citetalias{zhuravlev-razdoburdin-2014},
whereas the viscous part of equations contains terms emerging due to the non-zero perturbation of shear, which we denote by $N_{r,\varphi}$, terms emerging due 
to the non-zero perturbation of the velocity divergence, which we denote by $B_{r,\varphi}$, and, additionally, the advective terms $\propto v_r$ 
must be retained on a global scale. Note that the advective terms do not affect the dynamics of local perturbations, see comments 
after equations (\ref{sys_f}) and (\ref{sys_g}).
Explicitly,
\begin{equation}
\label{direct1}
\frac{\partial \delta v_r}{\partial t}+{\rm i}m\Omega \delta v_r=2\Omega \delta v_{\varphi} -\frac{\partial \delta h}{\partial r}-\frac{\partial}{\partial r}\left(v_r \delta v_r\right)+N_r+B_r,
\end{equation}

\begin{equation}
\label{direct2}
\frac{\partial \delta v_{\varphi}}{\partial t}+{\rm i}m\Omega \delta v_{\varphi}=-\frac{\kappa^2}{2\Omega}\delta v_r - \frac{im\delta h}{r}-\frac{v_r}{r}\frac{\partial}{\partial r}\left(r \delta v_{\varphi}\right)+N_{\varphi}+B_{\varphi},
\end{equation}

\begin{equation}
\label{direct3}
\frac{\partial \delta h}{\partial t}+{\rm i}m\Omega \delta h=-\frac{a_*^2}{\Sigma r}\frac{\partial}{\partial r}\left[r\Sigma\left(\delta v_r+\frac{v_r \delta h}{a_*^2}\right)\right]-\frac{{\rm i}m a_*^2}{r}\delta v_{\varphi},
\end{equation}
where the viscous terms $N_i$ and $B_i$ are shown in Appendix \ref{visc_app}.
In equations (\ref{direct1}-\ref{direct3}) we have also assumed that the Eulerian perturbations of dynamical and bulk viscosities are equal to zero 
$\delta \eta = \delta \zeta = 0 $. This is done to exclude viscous overstability of perturbations leading to their modal growth 
irrelevant to (both, modal and non-modal) amplification of non-axisymmetric perturbations, see e.g. \cite{kato-2001}.

\subsection{Adjoint equations}
\label{adj_eqs}

In order to obtain optimal perturbations via the iterative loop, we have to derive the set of adjoint equations, 
i.e. an explicit form of the adjoint operator $\mathbf{A}^{\dag}$ is needed.
In the global problem $\mathbf{A}^\dag$ is defined by the Lagrange rule (see equation (\ref{Lagr_rule}) ) through the inner product:
\begin{equation}
\begin{aligned}
&(\mathbf{q},\tilde{\mathbf{q}})=\pi\int\Sigma\bigg [\delta v_{r} \delta \tilde{v}_{r}^{*} + \left. \delta v_{\varphi} \delta \tilde{v}_{\varphi}^{*} + \frac{\delta h \delta \tilde{h}^{*}}{a_*^2}\right]rdr
\end{aligned}
\end{equation}
corresponding to a norm equal to total acoustic energy of perturbations
\begin{equation}
\label{glob_norm}
||\mathbf{q}||^2=E_a=\pi\int\Sigma\left(|\delta v_r|^2+|\delta v_{\varphi}|^2+\frac{|\delta h|^2|}{a_*^2}\right)rdr.
\end{equation}

Derivation of the adjoint equations is described in detail in \S 3.4 \citetalias{razdoburdin-zhuravlev-2015} for the inviscid case.
However, it can be straightforwardly generalised onto viscous problem.
We obtain:
\begin{equation}
\label{adjoint1}
\frac{\partial \delta \tilde{v}_r}{\partial t}+{\rm i}m\Omega \delta \tilde{v}_r=\frac{\kappa^2}{2\Omega}\delta \tilde{v}_{\varphi}-\frac{\partial \delta \tilde{h}}{\partial r}-\frac{v_r}{\Sigma r}\frac{\partial}{\partial r}\left(\Sigma r \delta \tilde{v}_r\right)+\tilde{N}_r+\tilde{B}_r,
\end{equation}

\begin{equation}
\label{adjoint2}
\frac{\partial \delta \tilde{v}_{\varphi}}{\partial t}+{\rm i}m\Omega \delta \tilde{v}_{\varphi}=-2\Omega \delta \tilde{v}_r-\frac{{\rm i}m\delta \tilde{h}}{r}-\frac{1}{\Sigma}\frac{\partial}{\partial r}\left(v_r \Sigma \delta \tilde{v}_{\varphi}\right)+\tilde{N}_{\varphi}+\tilde{B}_{\varphi},
\end{equation}

\begin{equation}
\label{adjoint3}
\frac{\partial \delta \tilde{h}}{\partial t}+{\rm i}m\Omega\delta \tilde{h}=-\frac{a_*^2}{r\Sigma}\frac{\partial}{\partial r}\left(r\Sigma \delta \tilde{v}_r\right)-\frac{{\rm i}m a_*^2}{r}\delta \tilde{v}_{\varphi}-v_r\frac{\partial \delta \tilde{h}}{\partial r}+\tilde{N}_h,
\end{equation}
where the viscous terms $\tilde N_i$ and $\tilde B_i$ are shown in Appendix \ref{visc_app}.

In order to solve the dynamical and the adjoint equations, we formulate the boundary conditions for viscous perturbations, see Appendix \ref{boundary_app}.

\subsection{Numerical method}
\label{num_meth}

For the numerical advance of perturbations obeying equations (\ref{direct1}) -- (\ref{direct3}) and (\ref{adjoint1}) -- (\ref{adjoint3}) we use the so called leap-frog method (see, e.g., \cite{frank-robertson-1988}).
This 2nd order method employs 4 uniform grids in the space $(r,t)$ shifted to each other, which allows to use central differences for the approximation of spatial derivatives of the unknown variables.
However, the approximation of the advection terms (that $\propto v_r$) by central differences leads to numerical instability.
In order to check this, one should apply Von Neumann stability analysis method, see \cite{charney-1950}.
That is why we use forward differences to approximate those terms in equations (\ref{direct1}) -- (\ref{direct3}), whereas in equations (\ref{adjoint1}) -- (\ref{adjoint3}) advanced back in time 
the advection terms are approximated by backward differences (a description of finite differences of various types can be found in \cite{olver-2014}).

Also note, that in order to evaluate $\partial \delta \tilde{h} / \partial t$ in the grid node neighbouring to the boundary, the value of $\delta \tilde{h}$ at the boundary is required since we use backward differences in the scheme 
for the adjoint equations. On the contrary, in the basic equations the forward differences are used, so the boundary condition for $\delta h$ is not required.

The allocation of the variables in the grids is the same as in the inviscid problem (see \citetalias[Fig. 10]{razdoburdin-zhuravlev-2015}).

\subsection{Results}
\label{glob_res}

As a result of the global optimisation problem we are going to obtain curves of optimal growth
\begin{equation}
\label{glob_G_t}
G(t) \equiv \max \frac{E_a(t)}{E_a(0)},
\end{equation}
where $E_a$ is given by equation (\ref{glob_norm}). 
Also, it is a common practice to consider another quantity
\begin{equation}
\label{glob_G_max}
G_{max} \equiv \max_{\forall t} G(t),
\end{equation}
which is similar to $G_{max}$ introduced by \citetalias{mukhopadhyay-2005} as well as \cite{yecko-2004} and \cite{maretzke-2014}, also cf. equation (\ref{G_max_compr}) as its local analogue.
Note that $G_{max}$ depends on the parameters of disc such as $\delta$, $R_{05}$ and $R_b$ and on azimuthal wavenumber $m$ of perturbations.
In this Section we present the results related to the first azimuthal harmonic of global perturbations, $m=1$, with polytropic index $n=3/2$. 
Also, to exclude the global analogue of density waves excitation during the swing of a global spiral of perturbations
(see \S \ref{bulk_visc} for details about the corresponding effect in the local framework), we use high value of bulk viscosity $R_b = 0.001 R_{05}$ in all calculations. 
It is implied that time is expressed in units of $\Omega^{-1}(r_i)$, the frequencies are expressed in units of $\Omega(r_i)$ and radial distance is expressed in units of $r_i$.

To be able to make comparisons with the corresponding local results on transient growth of large-scale vortices, we additionally introduce a global analogue of $k_y$
\begin{equation}
\label{local_k_y}
\bar{k}_y=\frac{m H}{(2n+1)^{1/2} \, r}
\end{equation}
and a global analogue of $R$
\begin{equation}
\label{local_Re}
\bar{R}=\frac{a_*^2}{\Omega \nu},
\end{equation}
cf. equation (\ref{reynolds}).
In general, $\bar k_y$ and $\bar R$ are functions of $r$. However, in the models N2 and P2 they are constant throughout the disc.
Obviously, the transient growth of the respective local perturbations is determined by values of those quantities at a certain radial distance.
Also, it makes sense to give an interpretation of some features of global transient growth with the help of (\ref{local_k_y}) and (\ref{local_Re}).

\subsubsection{Models N1 and N2}


\begin{figure}
\includegraphics[width=1.\linewidth]{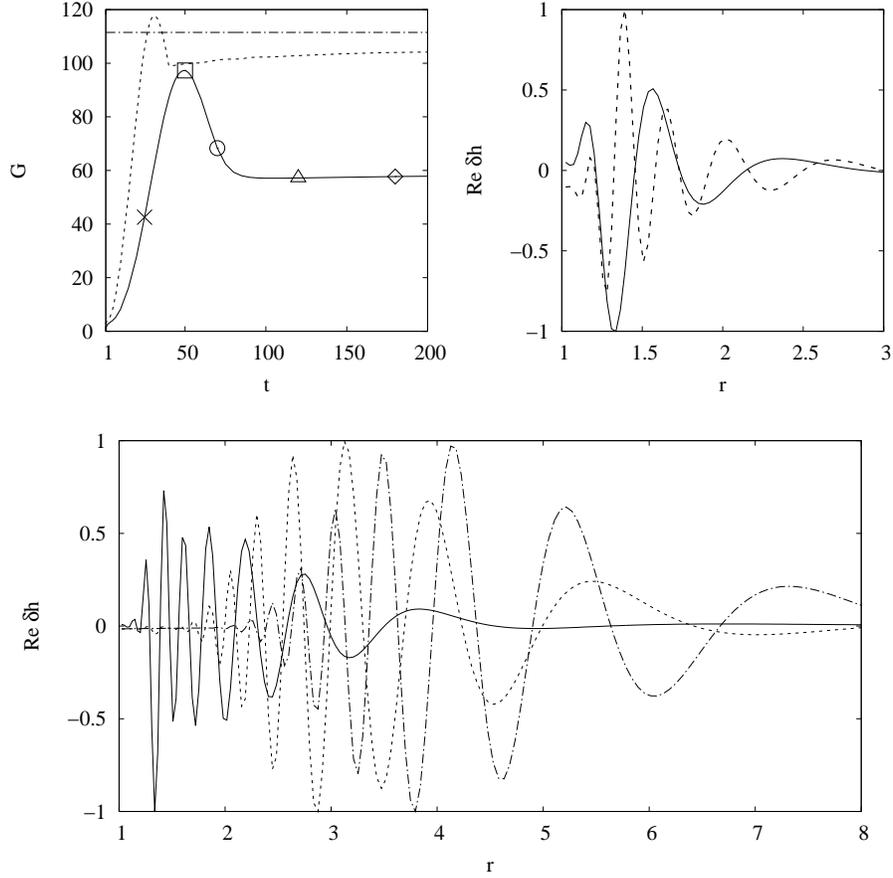}
\caption{
Optimal transient growth of global perturbations with $m=1$ in the model N2. Curves are presented for parameters $\delta=0.4$, $n=1.5$, $R_{05}=2000$ and $R_b=0.001 R$.
Top left panel (solid curve) shows the optimal growth as function of time, whereas top right and bottom panels show the profiles of $\Re~[\delta h]$ expressed in arbitrary units vs. the radial coordinate. 
These profiles correspond to the optimal initial perturbations.
Solid and dashed curves on top right panel correspond to the optimisation timespans denoted, respectively, by cross and square on top left panel.
Solid, dashed and dot-dashed curves on bottom panel correspond to the optimisation timespans denoted, respectively, by circle, triangle and diamond on top left panel.
Additionally, the dotted curve on top left panel shows the optimal growth for $m=10$ and $\delta=0.04$, whereas the dot-dashed curve is given by analytical estimation (\ref{anal_G_max_compr})
with $k_y=\bar k_y$ and $R=\bar R$ corresponding to the set of parameters used on this plot.
}
\label{Profiles_simple_Newton}
\end{figure}

Let us consider the optimal growth of global perturbations in the case of Newtonian background.
We start showing the behaviour of $G$ as a function of time in the model N2, see Fig. \ref{Profiles_simple_Newton}. 
At first glance, the optimal growth for $m=1$ still attains maximum which is expected given the results of local analysis, cf. bottom panel in Fig. \ref{fig_23}, though for some reason
$G$ tends to a non-zero value as $t\to \infty$. However, we additionally show the curve of $G(t)$ describing the behaviour of perturbations with $m=10$, which 
has to reproduce better the known bell-shape of local optimal growth curve obtained previously, e.g., by \citetalias{mukhopadhyay-2005}.
By contrast, it can be seen that as $m\to\infty$ the hump in $G(t)$ tends to disappear and the optimal growth almost monotonically approaches its plateau value across the considered time.
A clue to this discrepancy comes from checking profiles of optimal initial perturbations, see the other two panels in Fig. \ref{Profiles_simple_Newton}.
Clearly, the phase of growth of $G$ corresponds to the upwind of global leading spiral of initial perturbations. 
Using terms of the local approach, we say that the upwind continues until the time of order of $t_{max}$, see equation (\ref{t_max}), evaluated near the inner boundary of disc. Equation (\ref{t_max}) gives $t_{max}\approx 25$ for
the parameters used in Fig. \ref{Profiles_simple_Newton}, which is, indeed, close to what we find for perturbations with $m=10$.
As the characteristic time of viscous dissipation for spiral swinging at $t$  becomes shorter than $t$ (i.e. as $t>t_{max}$) its transient growth decreases, which is a single outcome for
local shearing harmonics as well as in the case of perturbations confined between the walls considered in previous studies. 
However, in the radially infinite flow similar to N2 the optimal spiral starts shifting outskirts instead of further upwind.
In terms of local approach this means that the instant of swing stays approximately equal to $t_{max}$ taken in units of $\Omega^{-1}$ at the new $r=r_{sp}$ where the spiral has shifted.
The time interval until the instant of swing, $t_{max} \Omega^{-1}(r_{sp})$, becomes longer and longer and it is always approximately equal to the optimisation timespan, $t$.
In fact, the latter is a general condition for the approximate localisation of global spiral:
\begin{equation}
\label{r_sp}
t_{max}(\,\bar R(r_{sp}), \bar k_y(r_{sp}),q(r_{sp})\,) \Omega^{-1}(r_{sp}) \approx t.
\end{equation}
At the same time, the global optimal spiral with $m\gg 1$ corresponding to $\bar k_y$ should recover the local value of $G$ in whatever part of disc it was localised.
But from the above it follows that this local value of $G$ must be equal to nothing but $G_{max}$. Then, since $\bar k_y$ and $\bar R$ (as well as $q$) are constant in N2, 
the locally determined $t_{max}$ and $G_{max}$ are also constant and for $m=10$ we do see the plateau of $G(t)$ at $t\gtrsim t_{max}\Omega^{-1}(r_i)$ which is 
close to the analytical value of $G_{max}$ given by equation (\ref{anal_G_max_compr}) with the values of $\bar k_y$ and $\bar R$ substituted in there, see the dot-dashed curve on top left panel in Fig. \ref{Profiles_simple_Newton}. 

The picture described above indicates that the hump in the curve of $G(t)$ and the related suppress of plateau for perturbations $m\sim 1$ in comparison with local value of $G$ are purely global features. 
Actually, this extra suppression of optimal growth takes place because of an additional spiral upwind after 
$G$ has already reached its maximum value corresponding to the equality of viscous dissipation time and optimisation time: compare the initial optimal profiles
corresponding to square and circle on the top left panel in Fig. \ref{Profiles_simple_Newton}. As has been discussed, such an additional upwind vanishes when $m\to\infty$. 
So, this purely global effect may be probably ascribed to a non-zero vorticity gradient since the most simple variant of background configuration corresponding to 
$\bar k_y=const$ and $\bar R=const$ has been considered herein. However, the clear explanation remains to be done.
Anyhow, as Fig. \ref{Profiles_simple_Newton} illustrates, the optimal spirals with $m=1$ also upwind/shift in a good agreement with the general picture that has been 
suggested and expressed through equation (\ref{r_sp}) just above: for example, equation (\ref{r_sp}) yields $r_{sp}\approx 3$ and $r_{sp}\approx 4$, respectively, for 
dashed and dot-dashed curves on bottom panel in Fig. \ref{Profiles_simple_Newton}.


\begin{figure}
\includegraphics[width=1.\linewidth]{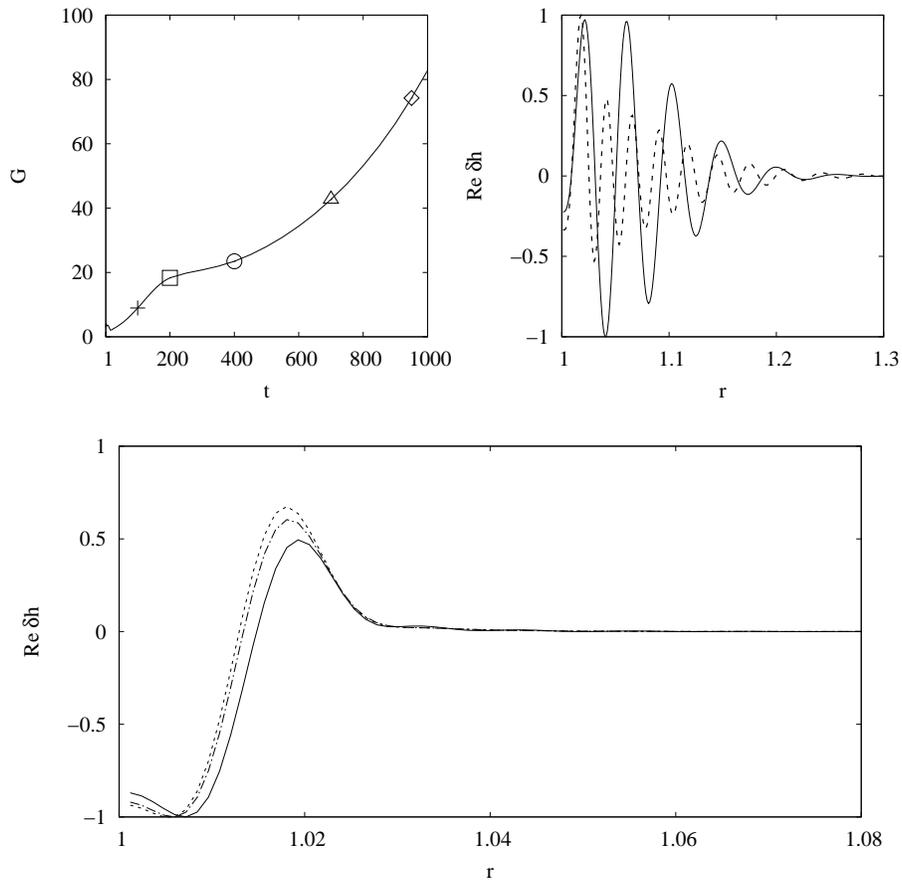}
\caption{
The same as in Fig. \ref{Profiles_simple_Newton} but for $\delta=0.02$.
Curves on top right and bottom panels correspond to the optimisation timespans marked by signs on top left panel 
in the same way as in Fig. \ref{Profiles_simple_Newton}.
}
\label{Profiles_simple_Newton_moda}
\end{figure}

Another purely global effect that is absent in the local formulation of the problem is modal growth of perturbations with $m\sim 1$ in the range of sufficiently small $\delta\lesssim 0.05$.
In Fig. \ref{Profiles_simple_Newton_moda} the dependence $G(T)$ and corresponding radial profiles of optimal initial perturbations are shown for $\delta=0.02$. 
In contrast to the previous case, a sufficiently thin disc represented by model N2 is unstable with respect to linear perturbations, which is indicated by $G$ exponentially growing 
at $t\to \infty$. Obviously, at large $t$ the optimal growth recovers the growth factor of the most unstable global mode. 
The amplitude of this mode increases $\propto \exp (\gamma t)$, with $\gamma\approx 0.001$ in Fig. \ref{Profiles_simple_Newton_moda}. 

Along with this fact, at small $t$ the optimal growth $G$ represents the value of transient growth, i.e. the transient growth dominates the modal growth.
This conclusion is drawn because the corresponding profiles of optimal initial perturbations are nothing but leading spirals, see the top right panel in Fig. \ref{Profiles_simple_Newton_moda}. 
Also, it can be checked that the growth factor of those spirals damps at large $t$.
Further, at longer $t$, see the bottom panel in Fig. \ref{Profiles_simple_Newton_moda}, the optimal initial profile has nothing in common with leading spirals and converges to the unique 
profile being the shape of the most unstable mode. We have checked that it corresponds to a pattern rigidly rotating with angular velocity $\omega_m\approx 0.98$, i.e. with 
corotation radius located inside of the flow. It is beyond the scope of this work to study the physical nature of the revealed instability, though presumably it is akin of the well-known 
Papaloizou-Pringle instability, see \cite{papaloizou-prinle-1987}, which has been also studied in details in 2D semi-infinite rotating flows by \cite{glatzel-1987a} and \cite{glatzel-1987b}.
Taking into account the results obtained in those papers we suppose that our unstable mode is a global non-axisymmetric sonic mode arising due to the existence of the boundary and supplied
with energy by the background flow at the critical layer localised at corotation radius. The latter mechanism of instability is usually called the Landau mechanism, see 
the book by \citet{stepanyants-fabrikant-1998} for the details. 
Our suggestion is confirmed by the fact that instability rapidly disappears as we go to perturbations with $m>1$, whereas the energy exchange between the mode and the flow is possible only 
in presence of the non-zero vorticity gradient which primarily affects perturbations on the largest spatial scale.


\begin{figure}
\includegraphics[width=1.\linewidth]{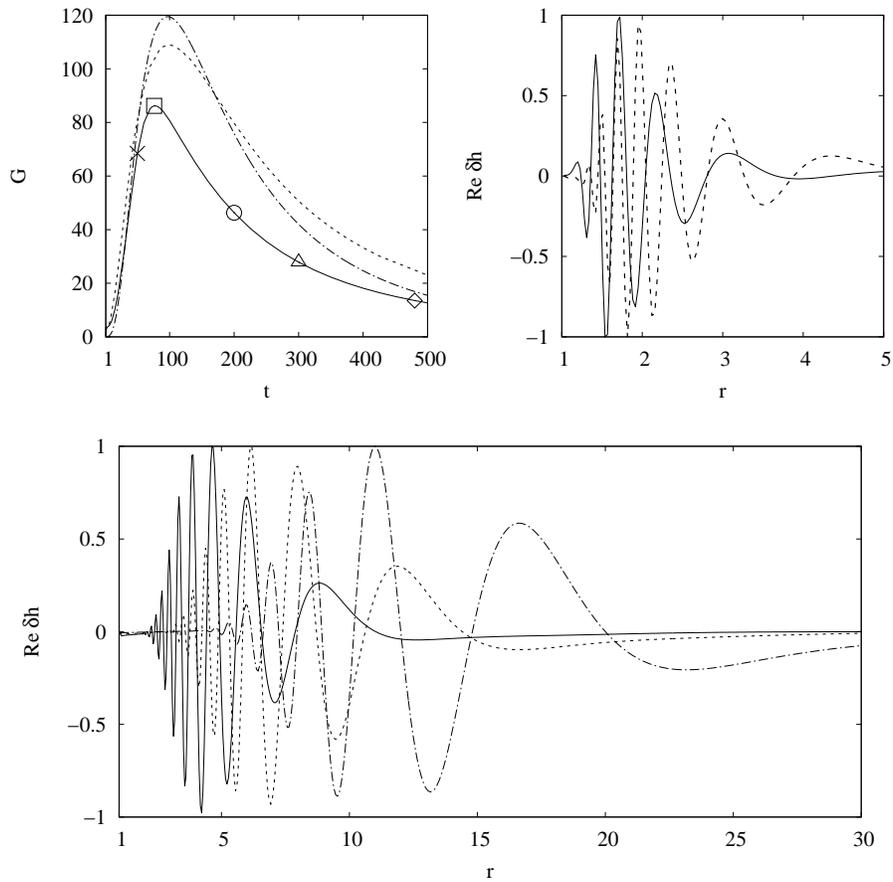}
\caption{
The same as in Fig. \ref{Profiles_simple_Newton} but in the model N1.
The curves on top right and bottom panels correspond to the optimisation timespans marked by signs on top left panel 
in the same way as in Fig. \ref{Profiles_simple_Newton}.
The dot-dashed curve on top left panel has been obtained using the local analytical estimations (\ref{t_max}) and (\ref{anal_G_max_compr}), see details in text.
}
\label{Profiles_real_Newton}
\end{figure}

Next, we proceed with model N1 keeping the same values of all parameters of the problem, see. Fig. \ref{Profiles_real_Newton}.
Clearly, the shape of $G(t)$ has substantially changed comparing to model N2. 
The plateau has disappeared and instead we obtain a curve with a global maximum and $G\to 0$ as $t\to\infty$ 
for perturbations with both $m=1$ and $m=10$. Yet, at the time after $G$ attains maximum this variant of $G(t)$ does not resemble the bell-shape curve known previously, see e.g. \citetalias{mukhopadhyay-2005}. Again, we see a complex behaviour of the initial optimal profile of perturbations combined of the spiral upwind and the spiral shift along the
radial direction. Curves on the bottom panel in Fig. \ref{Profiles_real_Newton} demonstrate that as the optimisation timespan increases, the initial optimal spiral
shifts outskirts and {\it unwinds}. This occurs because for $r>r_1$, see equation (\ref{nu_r}), the viscosity grows to the periphery of disc, 
so that $\bar R$ decreases to the periphery of disc for $r>r_1$. Consequently, the dimensionless $t_{max}$ expressed in units of $\Omega^{-1}(r)$ decreases outskirts as well. 
Further, as we have suggested above, the optimisation time is equal to the instant of swing of the optimal spiral which, in turn, is approximately equal to $t_{max}$. 
Since the degree of spiral winding, i.e. the initial ratio of radial and azimuthal wavenumbers  in terms of the local approach, 
is proportional to the dimensionless $t_{max}$ expressed in units of $\Omega^{-1}(r)$, we see that the optimal spiral must unwind as it shifts to the periphery of disc specified by the model N1. 
Note that $t_{max}$ expressed in units of $\Omega^{-1}(r_i)$, which is given by equation (\ref{r_sp}), remains to 
be a quantity which grows outwards, otherwise, the optimal spiral would not shifted at larger $r_{sp}$ for larger optimisation timespans.

In order to produce the analytical $G(t)$ corresponding to model N1, we fix an arbitrary $t$ and first estimate a location of global optimal spiral, $r_{sp}$, 
using the condition (\ref{r_sp}). After that, we find the respective value of $G_{max}(\, \bar k_y(r_{sp}), \bar R(r_{sp}), q(r_{sp}) \,)$.
This procedure allows us to obtain the dot-dashed curve on top left panel in Fig. \ref{Profiles_real_Newton} 
using the definitions (\ref{local_k_y}) and (\ref{local_Re}) together with specifications of the model given in \S \ref{basic_background}.
As we see, the analytical curve obtained by means of simple estimations of local optimal growth is in a reasonable agreement with the numerical one 
obtained employing the iterative loop for global perturbations with $m=10$. Once again, even the global spirals with $m=1$ upwind/unwind and shift along the radial 
direction in correspondence with interpretation produced in terms of local approach: for example, equation (\ref{r_sp}) yields $r_{sp}\approx 7$ and $r_{sp}\approx 12$, 
respectively, for dashed and dot-dashed curves on the bottom panel in Fig. \ref{Profiles_real_Newton}.

We have numerically checked, that the modal growth of perturbations with $m=1$ does not exist for any value of $\delta$ in the case of model N1.
Presumably, unstable mode which is located near the inner boundary of disc, see Fig. \ref{Profiles_simple_Newton_moda}, is suppressed by a sufficiently large viscosity 
therein, since in the model N1 $\bar R \to 0$ as $r\to r_i$.


\begin{figure}
\includegraphics[width=1.\linewidth]{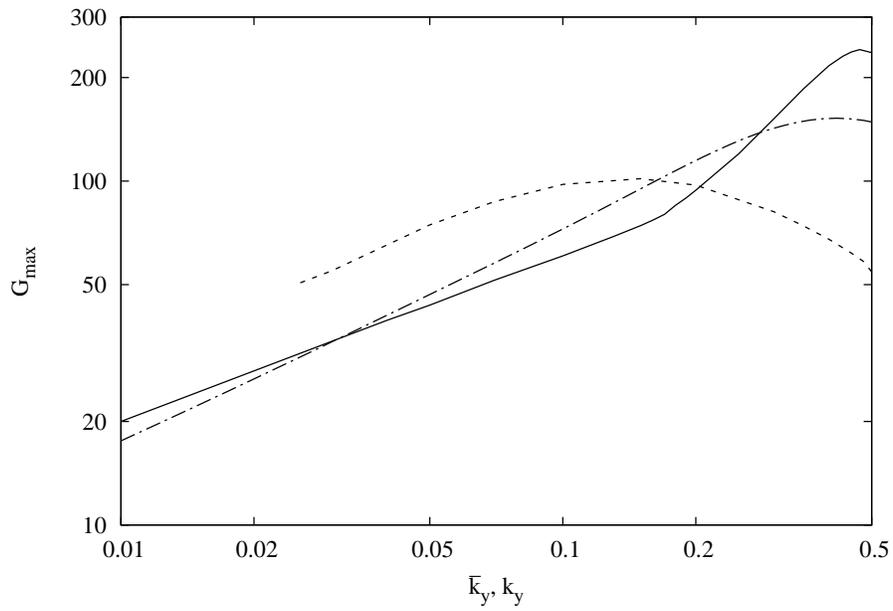}
\caption{
Dependence of $G_{max}$ on $\bar k_y\equiv m\delta/(2n+1)^{1/2}$ for $n=1.5$, $m=1$, $R_{05}=2000$ and $R_b=0.001R$.
Solid and dashed curves represent models N1 and N2, respectively, whereas dot-dashed curve represents $G_{max}(k_y)$ for local perturbations with $q=3/2$, $R_{05}=2000$ and $R_b=0.001R$. The termination of dashed curve corresponds to the occurrence of modal growth.
}
\label{G_k_y_Newton}
\end{figure}

To complete results related to the flow in the Newtonian potential, we compare the local dependence $G_{max}(k_y)$ with its global analogue, see Fig. \ref{G_k_y_Newton}.
In the case of global perturbations with $m=1$ $G_{max}$ can be defined as the height of hump in curve of $G(t)$. 
As can be seen in Fig. \ref{G_k_y_Newton}, both global and local large-scale vortices with $\bar k_y <1$ produce the transient growth of approximately the same 
value in discs with different aspect ratios and also in discs specified by different models. Moreover, the disc of the uniform surface density specified 
by model N2 gives rise to transiently growing spirals with $m=1$ which are able to exceed the local value of optimal growth as soon as $\bar k_y \lesssim 0.1$.  
At the same time, for larger values of $\bar k_y$ corresponding to geometrically thicker discs, as one considers the case of $m=1$ only, the model N2 
demonstrates that the global spirals exhibit a substantially weaker transient growth in comparison with the local value of $G$. 
The latter is in agreement with the results obtained in \citetalias{zhuravlev-razdoburdin-2014}, where in the inviscid problem it is revealed that the optimal 
growth of large-scale vortices (i.e. $\lambda_y > h$) stays almost at the same level as one passes from local perturbations to global perturbations 
with $m\sim 1$, whereas small-scale vortices (i.e. $\lambda_y <h$) with $m\sim 1$ reach substantially smaller growth factors than their local 
counterparts.

At last, let us make an important note about coexistence of modal and transient growth of perturbations with $m\sim 1$ in the sufficiently thin discs
specified by the model N2. In the Fig. \ref{Profiles_simple_Newton}-\ref{G_k_y_Newton} we fix $R_{05}=2000$ as it is done in the local problem for
easier comparison with the results of \citetalias{mukhopadhyay-2005}. However, one should keep in mind that actual Reynolds numbers in 
astrophysical discs are orders of magnitude larger and approach values like $\sim 10^{10}\div 10^{13}$. This implies that in practice 
the transient growth dominates the modal growth by many orders of magnitude for any value of $\delta$, since the optimal growth $\propto R_{05}^{2/3}$
whereas the increment increases by a few percent only as $R_{05}\to\infty$. 

\subsubsection{Models P1 and P2}


\begin{figure}
\includegraphics[width=1.\linewidth]{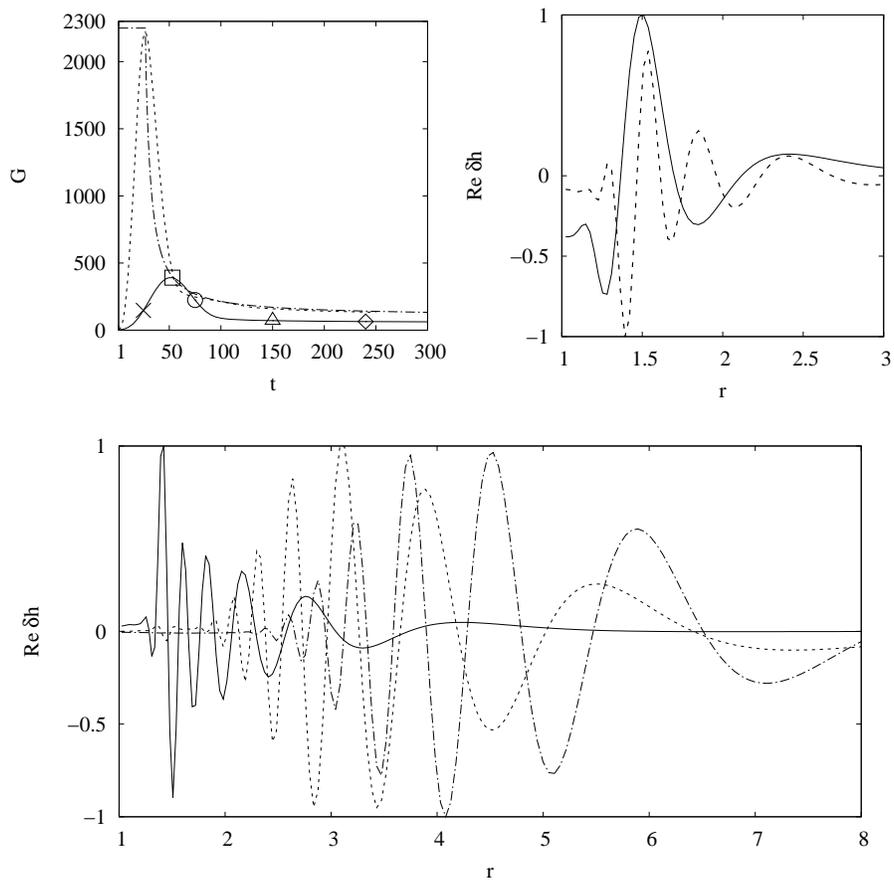}
\caption{
The same as in Fig. \ref{Profiles_simple_Newton} but in the model P2.
}
\label{Profiles_simple_Pachinsky}
\end{figure}

Let us see how the optimal growth changes as we consider the Paczynski-Wiita gravitational potential. In this case the epicyclic frequency (\ref{kappa2})
vanishes at the inner boundary, $\kappa(r_i)=0$, i.e. $r_i$ is the last stable circular orbit in the vicinity of gravitating centre.
According to the results obtained in the local approach, the optimal growth of large-scale perturbations significantly enhances as $q\to 2$.
As represented in Fig. \ref{fig_5}, for the case $k_y=0.2$ the local transient growth attains the value $g(t_s)\sim 10^3$, whereas at the same time
$G_{max}\approx 2250$ for $R_{05}=2000$ and $R_b\ll R_{05}$, see Fig. \ref{G_k_y_Pachinsky} below.
At first, we check the transient growth of global spirals with $m=1$ in the model of disc with uniform surface density which has been denoted as P2, see Fig. \ref{Profiles_simple_Pachinsky}. Following the strategy adopted to disc in the Newtonian potential, let us give an interpretation in terms of the local approach. 
Similar to the case of N2, the optimal initial spiral upwinds at short timespans while $t<t_{max}$, where $t_{max}\approx 20$ is 
estimated at $r_i$. After that, the spiral starts shifting outskirts which causes a decline of optimal growth. Now this decline of $G$ is much stronger 
than in the Newtonian case, cf. Fig. \ref{Profiles_simple_Newton}, since the value of optimal growth of large-scale vortices falls down as one approaches 
the Keplerian shear rate.
On the top left panel in Fig. \ref{Profiles_simple_Pachinsky} it is shown that for $t\gtrsim 20$ 
the analytical curve is in a moderate agreement with the numerical curve of $G(t)$ generated for global perturbations with $m=10$, and the agreement becomes
excellent when $t\to \infty$. The analytical curve corresponds to $G_{max}(\, \bar k_y=const, \bar R=const, q(r_{sp}) \,)$ given by equation (\ref{anal_G_max_compr}), where we substitute $r_{sp}$ evaluated from equation (\ref{r_sp}) into equation (\ref{q_r}). $G_{max}$ is not constant, indeed, due to the dependence $q$ on $r$ in P2. 
Also, since the analytical profile defined in this way diverges as $r_{sp}\to r_i$ we truncate it for $t\lesssim 20$ 
and replace by the numerical value of local $G_{max}=2250$ obtained for $k_y=0.2$ and parameters fixed in Fig. \ref{Profiles_simple_Pachinsky}. 
Note that the maximum of hump in $G(t)$ corresponding to $m=10$ is in a good agreement with the latest value.
Additionally, as $t\to\infty$ the optimal growth approaches its Newtonian values for both $m=1$ and $m=10$, cf. figs. \ref{Profiles_simple_Newton} and \ref{Profiles_simple_Pachinsky}. 

So, we see that the global perturbations with $m=1$ generate a much larger hump in P2 comparing to N2.
Obviously, this happens because the optimal large-scale vortex corresponding to short timespans is localised in the inner region of disc where $q$ is close to $2$.
Also, equation (\ref{r_sp}) yields $r_{sp}\approx 3$ and $r_{sp}\approx 4$, respectively, for dashed and dot-dashed curves on the bottom panel in Fig. \ref{Profiles_simple_Pachinsky}, 
which is again in a good agreement with the global numerical results obtained even for $m=1$.


\begin{figure}
\includegraphics[width=1.\linewidth]{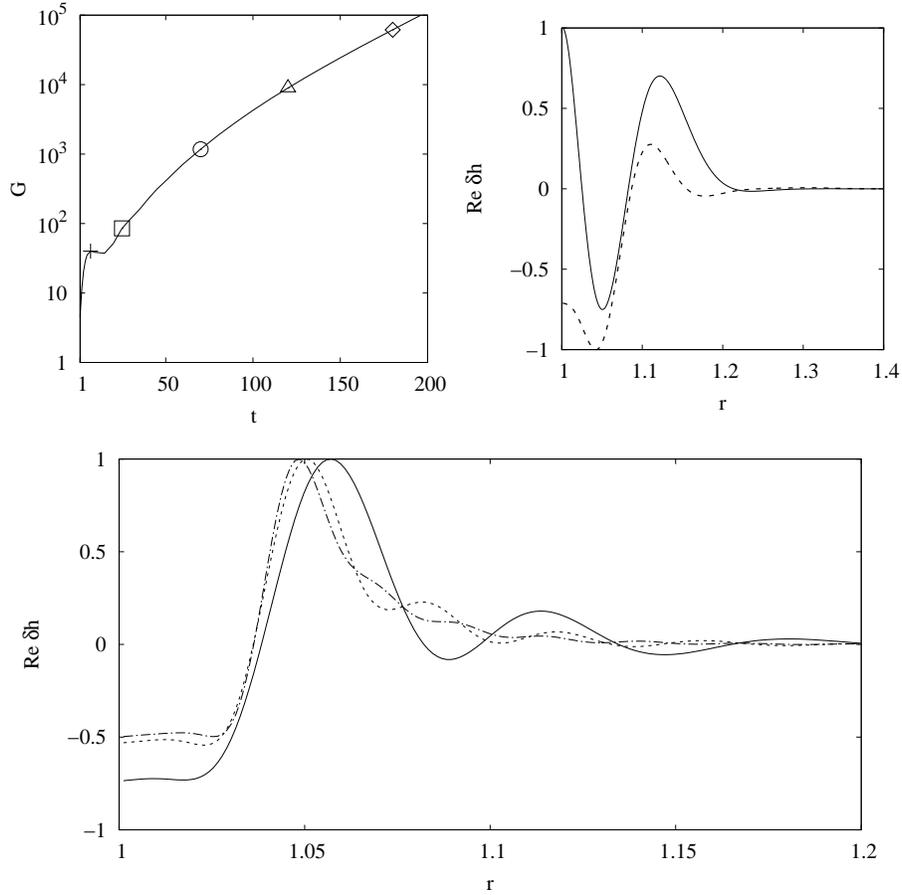}
\caption{
The same as in Fig. \ref{Profiles_simple_Newton_moda} but in the model P2.
}
\label{Profiles_simple_Pachinsky_moda}
\end{figure}

In the case of model P2, the modal growth emerges for $\delta<0.3$.
Similar to the model N2, the optimal perturbation at short time is a leading spiral exhibiting a transient growth of energy, see the top right panel
in Fig. \ref{Profiles_simple_Pachinsky_moda}. Then, for a sufficienly long $t$, we find that the optimal initial profile of 
perturbations converges to a rigidly rotating pattern, see bottom panel in Fig. \ref{Profiles_simple_Pachinsky_moda}.
We check that its increment has the maximum $\gamma \approx 0.03$ located at $\delta\approx 0.05$ for the particular values of free parameters chosen
to produce Fig. \ref{Profiles_simple_Pachinsky_moda}. 
The angular velocity of the unstable mode pattern represented in Fig. \ref{Profiles_simple_Pachinsky_moda} 
can be obtained via the Fourier decomposition of, say, $\Re [\delta h(t)]$ at some radius. We find that it corresponds to the corotation radius at $r_c \approx 1.04$, which is close to the inner boundary of disc.
Yet, the critical layer of mode is located inside the flow, so the necessary condition for modal energy transfer from background to perturbations is fulfilled. 
In spite of the fact that for $R_{05}=2000$ the instability exists in the substantial range of $\delta$ and the unstable mode has a noticeable increment, 
as one proceeds to larger Reynolds numbers typical for astrophysical discs, the transient growth starts dominating the modal growth at any $\delta$
for the same reasons as in the case of model N2.


\begin{figure}
\includegraphics[width=1.\linewidth]{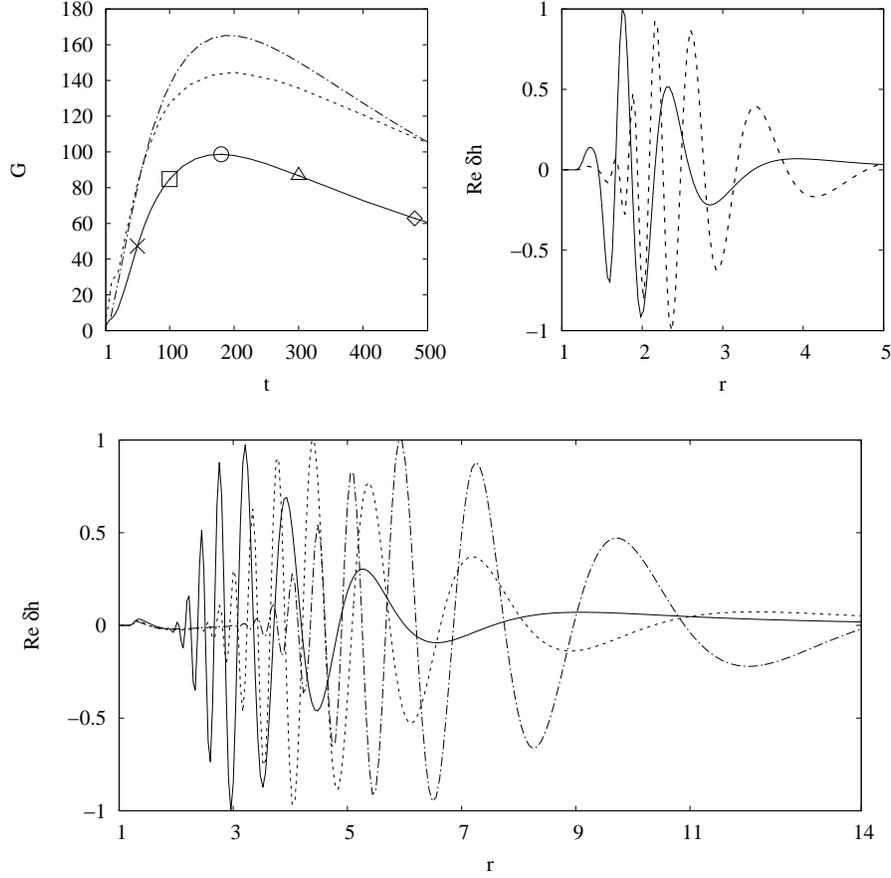}
\caption{ 
The same as in Fig. \ref{Profiles_real_Newton} but in the model P1.
}
\label{Profiles_real_Pachinsky}
\end{figure}

Finally, the numerical consideration of global perturbations reveals that at $R_{05}=2000$ the disc specified by the model P1 is linearly stable for all $\delta$, 
whereas the optimal growth is substantially suppressed in comparison with the model P2 and even more in comparison with the local $G_{max}$ for $q=2$
(see below the Fig. \ref{G_k_y_Pachinsky}). The curve of $G(t)$ and the behaviour of optimal initial spirals with $m=1$ can be seen in 
Fig. \ref{Profiles_real_Pachinsky}. Clearly, the optimal growth is suppressed in P1 because the optimal spiral corresponding to its maximum value 
is located somewhat farther from the inner boundary of disc than a similar spiral in P2, see the dashed curve on the top right panel in Fig. \ref{Profiles_simple_Pachinsky}
and the solid curve on the bottom panel in Fig. \ref{Profiles_real_Pachinsky}. In the former situation the spiral is centered at $r_{sp}\approx 1.5$, where $q\approx 1.8$,
while in the latter situation spiral is centered at $r_{sp}\approx 3$, where $q\approx 1.6$. For the local large-scale vortices such a difference in shear rate
corresponds to drop in transient growth by a factor of $\sim 3$. We see that for perturbations with $m=1$ it provides a somewhat larger difference in
the sizes of humps in $G(t)$ on top left panels in Fig. \ref{Profiles_simple_Pachinsky} and \ref{Profiles_real_Pachinsky}.
The reason for a spiral to shift out of the inner region of disc is that in P1 viscosity enhances as $r\to r_i$ which excludes the contribution of region with super-Keplerian 
shear rate close to $q=2$ to the transient growth of global perturbations.

Like in the other models considered before, we see a moderate agreement of $G(t)$ obtained for global perturbations
$m=10$ with its analytical counterpart defined by equation (\ref{anal_G_max_compr}) together with definitions (\ref{local_k_y}), (\ref{local_Re}), the condition (\ref{r_sp}) 
and specifications of the model given in \S \ref{basic_background}, see the top left panel in Fig. \ref{Profiles_real_Pachinsky}.
Additionally, equation (\ref{r_sp}) yields $r_{sp}\approx 4.5$ and $r_{sp}\approx 6.5$, respectively, for dashed and dot-dashed curves on the bottom panel in Fig. \ref{Profiles_real_Pachinsky}, which is also in agreement with global calculations for $m=1$.


\begin{figure}
\includegraphics[width=1.\linewidth]{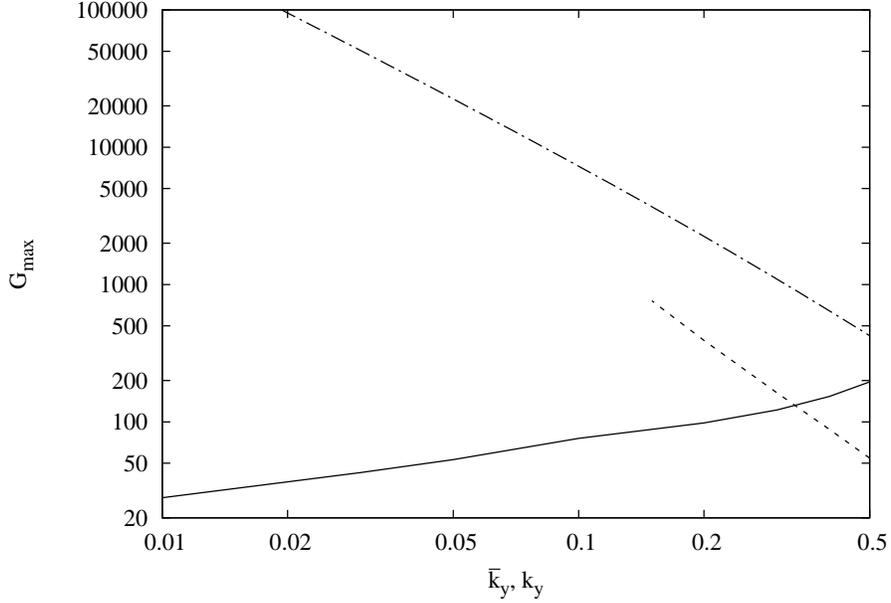}
\caption{
The same as in Fig. \ref{G_k_y_Newton} but
solid and dashed curves represent models P1 and P2, respectively, and dot-dashed curve represents $G_{max}(k_y)$ for local perturbations with $q=2$.
}
\label{G_k_y_Pachinsky}
\end{figure}

We finish the description of global optimal growth in the model P1 comparing its value for perturbations $m=1$ 
with the local analogue being $G_{max}(k_y)$ for $q=2$, see Fig. \ref{G_k_y_Pachinsky}.
In the case of global vortices we define $G_{max}$ as the height of hump in curve of $G(t)$, see the top left panel in Fig. \ref{Profiles_simple_Pachinsky}
and \ref{Profiles_real_Pachinsky}. 
For P2 $G_{max}$ is a few times smaller that local value, yet it raises up as $\bar k_y$ decreases similar to $G_{max}(k_y)$ until the modal growth
replaces the transient growth.
On the contrary, with regards to vortices with $m=1$, geometrically thin discs specified by the model P1 are subject neither to modal growth nor to 
transient growth which would exceed its Newtonian value.
Thus, in the model P1 the optimal growth of global perturbations, $G_{max}(\bar k_y)$, is similar to local optimal growth for the Newtonian shear rate
rather than for $q=2$.

\section{Conclusions}

We study the transient growth of shearing vortices at Keplerian and super-Keplerian shear rates taking into account a finite accretion disc thickness $H$. 
This allows us to consider perturbations of an arbitrary azimuthal length scale, $\lambda_y$, with respect to $H$ and, particularly, vortices with $\lambda_y \gtrsim H$ which we refer to as 'large-scale vortices' in contrast to small-scale vortices $\lambda_y\ll H$ received most of attention previously. 
In order to obtain reliable conclusions on their ability to exhibit a transient growth, we include viscous forces acting on fluid elements as a result of both their shear motions and their divergent motions. Thus, our dynamical model is parametrised not only by a Reynolds number, but additionally by a second Reynolds number corresponding to a non-zero bulk viscosity. 
At first, it is necessary to take into account the influence of bulk viscosity on the dynamics of shearing vortices, since close to the instant of swing a characteristic radial scale of perturbations is 
always larger than $H$ and velocity divergence becomes significant. This is of a special importance for large-scale vortices as their swing interval may be much longer than the rotational period of the flow.
Secondly, vortices with $\lambda_y\sim H$ generate shearing density waves that are primarily influenced by the bulk viscosity.

Mathematically, we identify the optimal perturbations exhibiting the highest possible transient growth among all perturbations of the same azimuthal wavelength. The value of maximum growth
factor is denoted by $G_{max}$. This is an important quantity usually determined in literature in the context of subcritical transition to turbulence in shear flows. 
Particularly, it has been demonstrated previously that in spectrally stable rotating shear flows $G_{max}$ correlates with the point of flow relaminarisation.
In order to obtain optimal solutions we employ an iterative loop constructed of the basic and adjoint equations advanced, respectively, forward and backward in time.  
Note that the optimisation performed via the iterative loop allows us to consider an unbounded shear flow.
In this work the adjoint equations have been derived for both local and global problems, see the sets (\ref{adj_SFH_sys_4}-\ref{adj_SFH_sys_6}) and (\ref{adjoint1}-\ref{adjoint3}), respectively.

We confirm that initial perturbations producing $G_{max}$ are vortical leading spirals, which is strictly shown by \citetalias{zhuravlev-razdoburdin-2014} for inviscid perturbations. 
In the large shearing box these spirals are represented by SFH with $k_x<0$. Being tightly wound, $|\tilde k_x|\gg 1$, they can be approximated by balanced solutions first used in \cite{heinemann-papaloizou-2009a}.
With the help of balanced solutions we derive an analytical estimate of $G_{max}$ for the large-scale vortices in viscous flow, see equation (\ref{anal_G_max_compr}).
It complements the known expression for $G_{max}$, see equation (\ref{anal_G_max}), which is valid for the small-scale vortices only, in the range of $\lambda_y\gg H$. 
Since equation (\ref{anal_G_max_compr}) yields $G_{max}\propto k_y^{2/3}$, there is no problem of divergence of optimal growth for long wavelength perturbations in the unbounded shear.
It is also important that the account of finite $H$ (and, thus, finite sound speed $a_*$) in the perturbed flow enables one to choose natural dimensionless variables in the problem, when 
$k_y$ is expressed in units of $H^{-1}$ rather than in units of an artificial $L^{-1}$, cf. Appendix \ref{subsect1} and \S \ref{subsect2}.

We start numerical analysis inspecting the influence of bulk viscosity on the growth of optimal SFH.
The results are shown in Fig. \ref{fig_23}, where one finds a sonic bump on the curves $G_{max}(k_y)$. As soon as the bulk viscosity becomes larger and larger, the sonic bump caused by the excitation of
shearing density waves shrinks and the profile of $G_{max}$ converges to the profile of $g_s(k_y)$ being the growth factor of the same optimal SFH evaluated 
at the instant of swing.
The suppression of excited density waves is illustrated in Fig. \ref{fig_4}.
Importantly, even a huge value of bulk viscosity does not affect the stage of transient growth as long as vortical spiral is being contracted by the shear, 
i.e. it does not cause damping of shearing vortex itself. 
Moreover, this holds throughout the whole range of azimuthal wavenumbers. Note that in Appendix \ref{bulk_app} we additionally give an analytical interpretation of this property of shearing vortices. 
Armed with this result, we choose $g_s$ to be a measure of transient growth of vortices in a compressible rotating flow. 
Obviously, $g_s$ is a quantity independent of a certain value of bulk viscosity.

It is shown that for $R_{05}=2000$ the growth of shearing vortices in the boundless Keplerian shear attains a maximum value $g_s\sim 100$ at $k_y\sim 0.3$ expressed in units of $H^{-1}$, which is substantially larger in comparison with corresponding $G_{max}\sim 20$ obtained by \citetalias{mukhopadhyay-2005} in an incompressible flow between the walls.
The optimal growth declines slower as $k_y\to 0$, rather than as $k_y\to \infty$, which is in agreement with analytical estimates (\ref{anal_G_max}) and (\ref{anal_G_max_compr}), see top panel in Fig. \ref{fig_23}.
In the case of a much larger Reynolds number typical for astrophysical discs 
turbulent energy reservoir can be, in principle, replenished by perturbations with $\lambda_y \gtrsim H$ 
provided that there is a subsequent delivery of their energy to SFH with negative $k_x<0$ on smaller azimuthal scales $\lambda_y<H$ by means of a non-linear interaction between SFHs. 
Of course, there must be a positive non-linear feedback for regeneration of large-scale leading spirals. 
\citet{shen-2006} performed high resolution 3D numerical simulations in the large shearing box approximation with horizontal size $L>H$, however, 
it was not still enough to well resolve the region $k_y<1$, where the transient growth appears to be the most prominent. 
Yet in the model of incompressible fluid \citet{lithwick-2009} had come to the conclusion that vortices of \citet{shen-2006} were not enough elongated in azimuthal direction to survive for longer times.
In light of our results, the high resolution simulations in box extended in azimuthal direction could resolve the tightly wound large-scale leading spirals as well as their possible repopulation due to the non-linear feedback
on the scale beyond the disc thickness.

At the same time, Fig. \ref{fig_5} demonstrates that approximately in the range $k_y<0.1$ in the Keplerian case $G_{max}$ is produced mostly at the expense                   
of enthalpy perturbation rather than velocity perturbation. It should be noted, however, that this perturbation of enthalpy has nothing in common with its akin existing in 
density waves: its amplitude increases as $\tilde k_x \to 0$ and, vice versa, decreases as $\tilde k_x$ of SFH starts growing back. 
Also, as it is suggested by balanced solutions, the perturbation of enthalpy entering the large-scale vortex is proportional to the potential vorticity perturbation.
By these reasons, it can be argued that even the large-scale SFH consisting mostly of $\hat W$ does not lead to generation of shocks but more likely 
reorganise a picture of non-linear interaction between SFHs, presumably, via the term $\nabla(\rho_1 {\bf u})$ in the continuity equation. 
For example, the last term must be significant if it is evaluated for density perturbation generated by swinging leading spiral with $k_y \ll 1$ and velocity perturbation 
generated by swinging leading spiral with $k_y \gtrsim 1$.

Considering super-Keplerian shear rates $1.5<q<2.0$, we find that the numerically obtained $g_s$ is also matched well by our analytical estimate of the optimal growth
(\ref{anal_G_max_compr}), see Fig. \ref{fig_5}. Thus, with the account of all perturbation scales as compared to $H$, 
it is confirmed that there is no such an abrupt depression of $G_{max}$, which takes place for perturbations $\lambda_y \ll H$ with the onset of stable epicyclic motions in the flow. 
For example, the shear rate $q=1.9$ provides a maximum growth of shearing vortices two orders of magnitude larger than it is found by \citetalias{mukhopadhyay-2005}
in incompressible flow. Notably, the fraction of kinetic energy, $g_{ks}/g_s$ increases as we approach $q=2$ for fixed $k_y$. This is shown also in Fig. \ref{fig_6},
where the amplitude of $\hat u_x$ taken at the instant of swing of optimal SFH raises faster than the amplitude of $\hat W$ as $q\to 2$. 
Additionally, as soon as $q\to 2$, maxima of $g_s(k_y)$ and $g_{ks}(k_y)$ shift to smaller $k_y$ and strictly for $q=2$ the growth factors diverge almost like $\propto k_y^{-4/3}$ exceeding the value of $G_{max}$ produced by streamwise streaks, cf. Fig. 2d by \citetalias{mukhopadhyay-2005}, at $k_y\lesssim 0.1$. At this point let us remark that \citet{rincon-2007} employed a sophisticated non-linear continuation methods to show rigorously that already in the case
of uniform angular momentum rotation, $q=2$, the non-linear feedback sustaining the linear transient growth of perturbations cannot be similar to the one described for plane shear flows and observed in weakly rotating cyclonic flow as well as
in Rayleigh-unstable anti-cyclonic flow. Partly, this is related to the fact that the finite-amplitude streamwise rolls generated by the anti-lift-up effect distort the rotational symmetry of the flow in sharp contrast to what takes place in the 
non-rotating lift-up case. At the same time, the anti-lift-up itself ceases to operate shortly after we shift to $q<2$. This suggests that, at least with the account of disc finite scale-height, the bypass mechanism of transition may
be based on the swing amplification of shearing spirals not only at super-Keplerian shear rate, but also in the uniform angular momentum rotation. 

The above results allow us to put a lower limit on the Reynolds number of subcritical transition to HD turbulence, see equation (\ref{R_q}). 
We suggest that the profile $R_T(q)$ may be followed by $\alpha(q)$, see equation (\ref{alpha}), where
$\alpha$ is the dimensionless azimuthal stress emerged from a hypothetical turbulence driven by growing shearing vortices. 
The dependence (\ref{alpha}) is steeper than obtained previously in simulations of supercritical MHD turbulence and, e.g., implies a substantially larger 
accretion rate corresponding to the Eddington luminosity in relativistic thin disc models. 

An azimuthal wavenumber of local perturbations cannot be as small as the disc aspect ratio, since $k_y \sim \delta$ would correspond to SFH with $\lambda_y \sim r_0$, 
which is not local anymore. Therefore, a study of transient growth of large-scale shearing vortices would not be complete without an investigation of 
global azimuthal harmonics of perturbations $\propto \exp({\rm i}m\varphi)$ with $m\sim 1$. In \S \ref{global} we additionally consider 
a global 2D dynamics of shearing vortices described by the set of equations (\ref{direct1}-\ref{direct3}) supplemented by global viscous terms
(\ref{direct4}-\ref{direct8}). We provide a detailed construction of several global background models, introduce boundary conditions
for both the basic and the adjoint variables and give brief notes on the numerical method. 
Note that here we leave out a problem of global density wave excitation setting $R_b\ll R$ in all global calculations. 
In contrast to the local vortex, the global one must be determined not only by the shear rate $q$ and the disc scale height $H$, 
but additionally by radial profiles of (i) a rotational frequency, (ii) a background vorticity, (iii) a surface density and a viscosity, (iv) 
a shear rate itself, e.g., in relativistic discs. The global optimal transient growth turns out to be more or less sensitive to all these factors, so
an exact shape of global $G_{max}(\bar k_y)$ depends on particular disc model. Nevertheless, we manage to understand the behaviour 
of global transient spirals with $m \gg 1$ employing interpretation in terms of the local approach and obtain a quantitative agreement between 
the shape of $G(t)$ for a variety of disc models and corresponding analytical estimations. The optimal spirals with $m\sim 1$ produce $G(t)$, which
deviates from the analytical curve. However, they basically preserve the qualitative properties of spirals with $m\gg 1$. 
It turns out that global optimal vortex is always radially localised and, as one
goes to longer timespans, the behaviour of its initial profile is a certain combination of the spiral upwind/unwind and spiral shift along the radial direction.
In particular, we manage to predict analytically locations of initial optimal vortices $m=1$ obtained for different optimisation timespans in any of the disc models we
used, see Fig. \ref{Profiles_simple_Newton}, \ref{Profiles_real_Newton}, \ref{Profiles_simple_Pachinsky} and \ref{Profiles_real_Pachinsky}.
In order to construct an analytical $G(t)$ we assume that (i) the optimisation time $t$ is equal to the time elapsed before the instant of swing $t_s$, (ii) as soon as $t$ becomes not less than $t_{max}$ evaluated at the inner boundary of disc $r_i$, it remains equal to $t_{max}$ evaluated at $r_{sp}>r_i$, 
where $r_{sp}$ is an approximate localisation of global spiral corresponding to this 
optimisation timespan. Note that assumption (i) is well justified for shearing vortices.
The assumptions (i) and (ii) allow us to formulate the condition (\ref{r_sp}), which provides us with estimation of $r_{sp}(t)$ and, subsequently, with estimation
of $G(t)$ which we suppose to be equal to local value of optimal growth at $r_{sp}$.
We find that optimal spirals corresponding to global $G_{max}(m=1)$ obtained in models with non-uniform viscosity and surface density profiles 
are shifted outskirts in comparison with similar spirals obtained in models with a constant viscosity and surface density. As soon as someone includes 
relativistic effects, cf. the models P1 and P2, this causes the suppression of $G_{max}$ down to Keplerian value, see Fig. \ref{G_k_y_Pachinsky}. 
At the same time, in the case of Newtonian dynamics we find the global $G_{max}$ to attain approximately the same value as its local analogue, cf. the
models N1 and N2 in Fig. \ref{G_k_y_Newton}. Consequently, the Keplerian value of optimal growth produced by
the large-scale shearing vortices is robust with the account of global effects, whereas the enhancement of locally determined $G_{max}$ obtained 
at super-Keplerian shear rate does not manifest itself on a global scale in a disc with viscosity attaining relatively high values in regions with
super-Keplerian rotation. Additionally, we consider the modal growth of global perturbations in sufficiently thin discs specified by the models 
N2 and P2. The most unstable mode is obtained by the optimisation method applied at long timespans. Its profile is shown is Fig. \ref{Profiles_simple_Newton_moda}
and Fig. \ref{Profiles_simple_Pachinsky_moda} for models N2 and P2, respectively. As can be seen, the unstable mode is concentrated close to the inner
boundary of disc. It is beyond the scope of this work to investigate the physical nature of modal growth, but 
we check that corotation of mode is located inside the flow and the increment attains a few percents larger value in the inviscid case. Also, the modal growth
ceases rapidly as we increase $m$. Thus, we suspect that the revealed instability is explained by the Landau mechanism implying that the 
energy flux from the background flow to the acoustic mode is caused by a non-zero gradient of background vorticity at corotation point.
Let us point out that this sort of modal growth competes with transient growth at Reynolds numbers equal to few thousands only. 
Since $G_{max}\propto R^{2/3}$, we expect that in astrophysical situation $R\sim 10^{10}\div 10^{13}$
the transient growth of global perturbations dominates the modal growth in a disc of an arbitrary thickness.

\subsection{Future prospects}
\label{future}

This work has demonstrated the importance of considering the transient growth at all scales with respect to the disc scaleheight. 
However, it has been carried out in a simplified 2D approximation of perturbation dynamics.
Accordingly, the optimal vortices should be subsequently studied with the account of vertical motions. 
It is worth checking whether 3D optimal perturbations with either $\lambda_y\ll H$ or $\lambda_y\gtrsim H$ in {\it radially unbounded} flow have a non-trivial axial structure and 
kinetic energy transferring into vertical motions in connection with the results obtained by \citet{maretzke-2014}. 
Indeed, within the model of fluid between the rotating cylinders, \citet{maretzke-2014} have compared the optimal growth in various regimes of a spectrally stable incompressible Taylor-Couette flow. 
They found that in contrast to the quasi-Keplerian regime, the cyclonic and the counter-rotating regimes allow for the axial dependence of optimal 
vortices as well as for a significant part of the kinetic energy to be contained in vertical motion. Though physical reasons for such a difference remain unknown, 
this can probably be related to a subcritical HD transition in cyclonic and counter-rotating regimes occurring at ordinary values of Reynolds number contrary 
to the case of quasi-Keplerian regime. Thus, an interesting issue would be to find out whether the optimal\footnote{or, {\it nearly} optimal} perturbations in an unbounded quasi-Keplerian flow
have the non-trivial axial dependence and to explain its physical nature. 
Note that with regards to the large-scale vortices, $\lambda_y\gtrsim H$, the corresponding result may be different as one formally considers the vertically homogeneous or more realistic vertically baratropic flow.

While there is expectation to reveal the bypass transition within the most basic model of quasi-Keplerian flow, astrophysical discs can be stratified and 
there is a number of refined models shown to exhibit linear and non-linear HD activity due to various thermodynamical inhomogeneities. As a rule, this HD activity results in generation of
vortices of considerable amplitude, which are usually invoked to resolve the problems of angular momentum transport and planetesimal formation inside non-ionised parts of protoplanetary discs. 
Some of them are the product of modal instabilities, such as Rossby wave instability \citep{lovelace-1999,lovelace-2000,lovelace-2001} or vertical shear instability \citep{urpin-2003,nelson-2013,richard-2016}, 
whereas the others are attributed to subcritical instabilities emerging either due to the radial entropy gradient \citep{lesur-papaloizou-2010}, or due to the vertical entropy gradient \citep{marcus-2015}.
In view of subcritical behaviour of stratified Keplerian flow it is important to note that a novel type of linearly growing vortical 3D perturbations appears due to the so called 
strato-rotational balance, see \citet{tevzadze-2003} and \citet{tevzadze-2008}. The existence of these 3D vortices is closely related to special invariant of stratified fluid dynamics
equal to scalar product of the entropy gradient and the vortensity. These vortices have been suggested as transient perturbations for bypass transition scenario in stratified discs, however, note
that they degenerate into trivial solution within unstratified model considered in this paper. Also interestingly, that the presence of radial stratification allows for the production of vortices
by means of purely linear dynamics, see \citet{tevzadze-2010}. The 2D shearing density waves acquire the ability to generate 2D shearing vortices through indirect coupling with the assistance of shearing entropy waves. 
It is not yet clear what role this novel and complicated non-modal linear dynamics may play in subcritical instability of stratified Keplerian flows, motivating to consider the corresponding 3D optimisation problem for
perturbations with the account of the finite disc thickness.  

Furthermore, it seems relevant to study the excitation of density waves within the 3D optimisation problem in baratropic and vertically stratified backgrounds. Particularly, this
would be necessary to do in the large-scale limit $\lambda\gg H$ in order to establish the link with the work done by \citet{umurhan-2006} and later \citet{rebusco-2009}. 
Employing the asymptotic time-dependent model of geometrically thin viscous Keplerian disc, these authors obtained that the vertical sound waves exhibit non-modal growth being coupled to planar epicyclic oscillations.

At last, it is desirable to perform high-resolution HD simulations in the large shearing box approximation in order to check whether the critical value of the shear rate $q<2$
corresponding to turbulence decay found by \citet{hawley-1999} gets smaller as one considers the non-linear dynamics of shearing vortices with $k_y\lesssim 1$.
It would be particularly interesting to address the issue of significant density perturbations produced by large-scale vortices 
in the context of non-linear interaction between SFH with different $k_x$ and $k_y$. We point out that these density perturbations are 
proportional to a perturbation of the potential vorticity being time-independent in 2D inviscid compressible dynamics, so they are qualitatively
different from the density perturbations generated by shearing density waves. Also, it is of course an open question in which way these density perturbations 
would affect a hypothetical non-linear transverse cascade which probably provides a positive feedback for transiently growing leading spirals at
very high Reynolds numbers in the quasi-Keplerian flows. 

\subsection{Final remarks}

Let us note that the issues we have mentioned in this paper are closely related to the applied problem of stability and transition to turbulence in hypersonic shear flows. 
The stability and transition in hypersonic shear flows is a developing area of experimental, theoretical and numerical fluid dynamics with a particular emphasis on the design of flying vehicles, see
the book by \citet{gatski-bonnet-2009} and the reviews by \citet{fedorov-2011} and \citet{zhong-wang-2012}. These flows are the usual phenomena in space
and should be understood better in astrophysical science and, mainly, in theory of astrophysical discs. 

We would like to finish this paper making link with a neighboring field of magnetohydrodynamic (MHD) turbulence in accretion discs.
Actually, it became obvious in the past few years that the bypass concept of the transition to turbulence and mechanisms of transient growth are relevant not only to the HD problem.
The work by \citet{squire-bhattacharjee-2014b} and \citet{squire-bhattacharjee-2014} (see also \citet{nath-2015}) has shown that the established practice to oppose the subcritical HD transition to supercritical 
magnetohydrodynamic (MHD) transition to turbulence in astrophysical discs was misleading in a particular sense. It turned out that even in a spectrally unstable magnetised Keplerian shear the 
transient growth prevails the modal growth by orders of magnitude at certain short timespans. Thus, in fact, it is not yet clear what linear mechanism gives rise to a sustained MHD turbulence even in hot 
ionised accretion discs: either the magnetorotational instability, or the transient growth of magnetised shearing vortices. 
Moreover, a non-modal amplification of the magnetic field and subsequent positive non-linear feedback analogous to SSP in the context of HD
transition in plane shear flows has been suggested by \citet{rincon-2008} and \citet{riols-2013} to give a basic explanation of the subcritical magnetorotational dynamo action. The latter is crucial for
turbulisation of spectrally stable Keplerian background with zero net magnetic flux. 
The above extension of non-modal concepts onto MHD turbulence in astrophysical flows should be completed by the subcritical transition to turbulence via the non-linear transverse cascade providing positive feedback 
to transiently growing SFH in 2D magnetised plane flow, recently demonstrated by \citet{mamatsashvili-2014}.
Bearing in mind the current situation in the problem of the transition to turbulence in (quasi-)Keplerian flow, it seems to be fruitful to address remaining issues in both HD and MHD problems using a general non-modal approach.

\section*{Acknowledgements}

We thank P. B. Ivanov for careful reading the manuscript and important remarks. 
DNR was supported by grant RSF 14-12-00146 for writing \S \ref{global} of this paper.
VVZ was supported by RFBR grant 15-02-08476 А.
Also, the study was in part supported by the M. V. Lomonosov Moscow State University Programme of Development
and in part by the Programme 7 of the Presidium of the Russian Academy of Science.

\appendix

\section{Subcase of incompressible dynamics}
\label{incompr}
\subsection{Shearing vortex solution}
\label{subsect1}

Let us consider the incompressible limit of transient dynamics also studied by \citet{yecko-2004} and \citetalias{mukhopadhyay-2005} 
in the case of a flow confined between two walls separated from one another by distance $2L$ across the shear.
On the contrary, here we are going to deal with an unbounded shear. 
As $a_*\to \infty$, the continuity equation (\ref{sys3}) yields the following restriction on the velocity components 
$$
\frac{\partial u_x}{\partial x} = - \frac{\partial u_y}{\partial y},
$$
and we are left with two projections of Euler equation (\ref{sys1}-\ref{sys2}). 

We change to dimensionless comoving Cartesian coordinates in (\ref{sys1}-\ref{sys3}),
\begin{equation}
\label{incompr_coords}
x^\prime =  x/L,\, y^\prime = (y+q\Omega_0 xt)/L,\, t^\prime=\Omega_0 t,
\end{equation}
where the quantity $L$ is an auxiliary radial scale, which does not have any physical meaning. 
The coordinates (\ref{incompr_coords}) lead us to a spatially homogeneous set of equations which have partial solutions in the form
of SFH corresponding to the dimensionless wavenumbers $k_x$ and $k_y$ expressed in units of arbitrary $L^{-1}$.
Similarly to \S\ref{subsect2}, we omit the primes and introduce the shearing radial wavenumber $\tilde k_x \equiv k_x + q k_y t$ and
the full wavenumber squared $k^2 \equiv \tilde k_x^2 +k_y^2$, arriving at the following dimensionless equations for a single SFH:

\begin{equation}
\label{SFH_sys_1}
\frac{d \hat u_x}{d t} = 2\hat u_y - {\rm i}\tilde k_x \hat W - R^{-1} k^2 \hat u_x,
\end{equation}

\begin{equation}
\label{SFH_sys_2}
\frac{d \hat u_y}{d t} = -(2-q)\hat u_x - {\rm i}\tilde k_y \hat W - R^{-1} k^2 \hat u_y,
\end{equation}
provided that
\begin{equation}
\label{SFH_sys_3}
\tilde k_x \hat u_x = -k_y \hat u_y,
\end{equation}
and it is assumed that $\hat u_x,\hat u_y$ are expressed in units of $U \equiv \Omega_0 L$. $\hat W$ is expressed in units of $U^2$. 
Correspondingly, $R \equiv U L / \nu$ is Reynolds number. Note that \citetalias{mukhopadhyay-2005} used Reynolds number 
\begin{equation}
\label{R_05}
R_{05} = q R.
\end{equation}

Using (\ref{SFH_sys_1}-\ref{SFH_sys_3}), we obtain equation for $\hat u_x$ which reads
\begin{equation}
\label{Eq_incompr}
\frac{d\hat u_x}{dt} = -2q \frac{k_y \tilde k_x}{k^2} \hat u_x -R^{-1} k^2 \hat u_x.
\end{equation}

It can be checked that (\ref{Eq_incompr}) is reproduced from equation (16) of \citetalias{mukhopadhyay-2005} and has the following analytical solution:
\begin{equation}
\label{sol_incompr}
\hat u_x (t) = \hat u_x(0) \frac{k_x^2+k_y^2}{k^2} \exp\left ( -\frac{\tilde k^3-k_x^3}{3qk_y R} - \frac{k_y^2 t}{R} \right ).
\end{equation}

The solution (\ref{sol_incompr}) is used to obtain the transient growth factor of incompressible perturbations for particular values of $k_x, k_y$, $R$ and $q$, see \S \ref{subsect3} for further consideration.

\subsection{Transient growth factor}
\label{subsect3}

As usual, the size of incompressible perturbations in an unbounded shear flow is measured by a surface density of their kinetic energy, $E_k$, which leads to simple expression for a single SFH:
\begin{equation}
\label{incompr_norm}
E_k = \frac{1}{2} ( |\hat u_x|^2 + |\hat u_y|^2) = \frac{1}{2} \frac{k^2}{k_y^2} |\hat u_x|^2.
\end{equation}
Substituting equation (\ref{sol_incompr}) into (\ref{incompr_norm}) yields the following growth factor of SFH:
\begin{equation}
\label{g_incompr}
g \equiv \frac{E_k(t)}{E_k(0)} = \frac{k_x^2+k_y^2}{k^2} \exp\left ( -2\frac{\tilde k^3-k_x^3}{3qk_y R} - 2\frac{k_y^2 t}{R} \right ).
\end{equation}

Given the particular values of $k_y$, $R$ and $q$, (\ref{g_incompr}) has a global maximum over all $k_x$ and $t$ which we are going to denote here as
\begin{equation}
\label{G_max_incompr}
G_{max} \equiv \max_{\forall k_x, \forall t}\, g,
\end{equation}
since this quantity is analogous to the maximum optimal growth defined in \S \ref{subsect4}, see equation (\ref{G_max_compr}).

\section{Analytical estimations\\ of optimal growth}
\label{prelim}

\begin{figure}
\begin{center}
\includegraphics[width=9cm,angle=0]{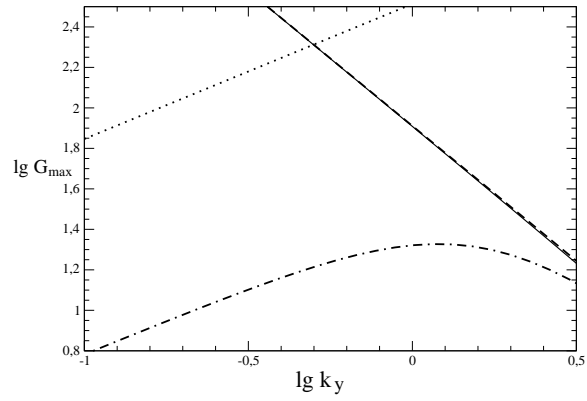}
\end{center}
\caption{Curves of $G_{max}$ in the subcase of incompressible dynamics of 2D perturbations for $R_{05}=2000$ and $q=3/2$: 
solid curve is obtained numerically using the definition (\ref{G_max_incompr}), dashed line represents equation (\ref{anal_G_max}), 
dot-dashed curve corresponds to equation (77) of \citetalias{mukhopadhyay-2005} taken with $k_{X,min}L=1.7$ in their notations. 
The dotted curve represents approximate $G_{max}$ given by equation (\ref{anal_G_max_compr}) for the case of finite disc thickness and
$k_y$ normalised by $H$ rather than by $L$.} \label{fig_1}
\end{figure}

By means of the analytical supplement to their numerical optimisation results, \citetalias{mukhopadhyay-2005} demonstrate that in the case of incompressible dynamics $G_{max}(k_y)$ 
acquires a maximum owing to the presence of rigid boundaries (see also the results of \citet{yecko-2004} in its Fig. 3b). 
In Fig. \ref{fig_1} we take the Keplerian shear, $q=3/2$, $R_{05}=2000$ and plot equation (77) of \citetalias{mukhopadhyay-2005} (dot-dashed line)
together with our quantity (\ref{G_max_incompr}) obtained numerically (solid line). It can also be checked that equation (77) of \citetalias{mukhopadhyay-2005} taken for a boundless shear, 
\begin{equation}
\label{anal_G_max}
G_{max} = \left ( \frac{qR}{k_y^2} \right )^{2/3} \, {\rm e}^{-2/3},
\end{equation}
virtually coincides with our numerical curve in Fig. \ref{fig_1}. Equation (\ref{anal_G_max}) can be easily obtained from (\ref{g_incompr}) setting $\tilde k_x=0$ (the swing instant of SFH) and taking the limit $|k_x| \gg k_y$.
Thus, in general, $G_{max}\to \infty$ as $k_y\to 0$ in the absence of boundaries. Physically, this takes place due to a simple fact that for a tightly wound SFH with $|k_x|\gg k_y$ 
both viscous dissipation time-scale, $t_\nu$, and transient growth duration, $t_{tg}$, 
become longer as we proceed to small azimuthal wavelengths, whereas the (inviscid) growth factor increases monotonically with time, see equation (\ref{g_incompr}).  
As it is discussed, e.g., by \citetalias{razdoburdin-zhuravlev-2015}, equation (\ref{anal_G_max}) can be approximately derived suggesting that $G_{max}$ corresponds to equality $t_\nu=t_{tg}$, which gives the time when the maximum growth occurs as
\begin{equation}
\label{t_max}
t_{max} \approx R^{1/3} (qk_y)^{-2/3}.
\end{equation}
Note that estimation (\ref{t_max}) provides exactly the same value which one gets looking for a maximum of (\ref{g_incompr}) in order to derive an approximate $G_{max}$ given by equation (\ref{anal_G_max}).
Then, using the inviscid form of $g\approx (qt)^2$, we recover equation (\ref{anal_G_max}) up to the damping factor $\exp(-2/3)$.

Of course, the growth factor (\ref{g_incompr}) and so $G_{max}$ do not depend on the scaling factor $L$.
As we see from equation (\ref{anal_G_max}), the transient growth value is controlled by the ratio $R/k_y^2 \sim (\lambda_y/l_p)^2$, where $\lambda_y$ is characteristic azimuthal length of perturbations and $l_p$ is 
mean free path of gas particles which determine microscopic viscosity. 
In fact, $\lambda_y^{-1} \sim k_y L^{-1}$ of SFH exhibiting $G_{max}$ cannot be too low since $|k_x|\sim (qk_y R)^{1/3}$ also decreases: but $|k_x| L^{-1}$ cannot be less than $H^{-1}$, 
since in the last case we start dealing with patch of disc comparable to its thickness in the direction of shear what forces us to take into account the effects of compressibility, 
see e.g. the discussion about SFH swing interval in \S 3.2.1 of \citetalias{zhuravlev-razdoburdin-2014}. Thus, we of course cannot reach an arbitrary high $G_{max}$ at the expense of $\lambda_y$ value.
It turns out, see \S 2.3 of \citetalias{razdoburdin-zhuravlev-2015}, that changing to a model with the account of finite disc thickness, we can 
derive an approximate expression for $G_{max}$ in the opposite case of $\lambda_y\gg H$ (i.e. $ k_y\ll 1$) employing the balanced solutions to equations (\ref{SFH_sys_4}-\ref{SFH_sys_6}), see \citet{heinemann-papaloizou-2009a}.
The inviscid result reads $g_{opt} \approx 4\kappa^{-4}(k_y q t)^2$, where $\kappa^2=2(2-q)$ is the dimensionless epicyclic frequency squared expressed in units of $\Omega_0^2$.
Substituting $t_{max}$ given by equation (\ref{t_max}) into this result we obtain
\begin{equation}
\label{anal_G_max_compr}
G_{max} \approx \frac{1}{(2-q)^2} (k_y q R)^{2/3} \, {\rm e}^{-2/3},
\end{equation}
where damping factor $\exp(-2/3)$ has been added since we expect a similar action of viscous forces on shearing vortices having both $k_y\gg 1$ and $k_y \ll 1$, see a detailed justification of this assumption in \S \ref{bulk_visc}.
As we see, one gets an opposite dependence of the maximum optimal growth on $k_y$, i.e. $G_{max}$ decreases as we approach larger azimuthal scales. 
Thus, there is no problem of divergence of transient growth factor towards $k_y \to 0$, see Fig. \ref{fig_1}.
At the same time, as it is seen in Fig. {\ref{fig_1}}, one expects much larger transient growth of columnar perturbations (i.e. perturbations independent of vertical direction) in the domain
of $\lambda_y \gtrsim H$, than it might seem within the model of \citet{yecko-2004} and \citetalias{mukhopadhyay-2005}.

\section{Suppression of shearing density waves for tightly wound spirals}
\label{bulk_app}

In the limit $|\tilde k_x|\gg 1$\footnote{or, equivalently, outside of the swing interval of SFH, see \citetalias{zhuravlev-razdoburdin-2014}.} the inviscid shearing vortices are approximately given by balanced solutions 
obtained from wave equations for $\hat u_{x,y}$ and $\hat W$, see e.g. equations (27-29) of \citetalias{razdoburdin-zhuravlev-2015}. Their explicit form is given by equations (35-37) of \citetalias{razdoburdin-zhuravlev-2015}. The non-zero viscosity leads to the following wave equations for $\hat u_y$ and $\hat u_x$:

\begin{multline}
\label{wave_visc_u_x}
\ddot{\hat u}_x = - k_y I_\nu -  K \hat u_x - 2{\rm i}qk_y \hat W  - \\ R^{-1} k^2 \left [  \dot{\hat u}_x + \hat u_y + 2q \frac{\tilde k_x k_y}{k^2} \hat u_x \right ] - \\ 
(R^{-1}/3 + R_b^{-1}) \left [ \frac{d}{dt}\left (\tilde k_x \, {\rm i} \dot{\hat W} \right ) + 2k_y \, {\rm i} \dot{\hat W} \right ],
\end{multline}

\begin{multline}
\label{wave_visc_u_y}
\ddot{\hat u}_y = \tilde k_x I_\nu -  K \hat u_y  - \\ R^{-1} k^2 \left [  \dot{\hat u}_y - (2-q)\hat u_x + 2q \frac{\tilde k_x k_y}{k^2} \hat u_y \right ] - \\ 
(R^{-1}/3 + R_b^{-1}) [k_y \, {\rm i} \ddot{\hat W} - (2-q)\tilde k_x \, {\rm i} \dot{\hat W}],
\end{multline}
where $K\equiv k^2 + \kappa^2$ and $I_\nu\equiv \tilde k_x \hat u_y - k_y \hat u_x + {\rm i}(2-q)\hat W$ is SFH of the potential vorticity perturbation, see corresponding expressions e.g. of \citet{bodo-2005} or \citet{tevzadze-2003} for the case of baratropic 
perturbations.
$I_\nu$ does not depend on time in the inviscid limit, however, in presence of non-zero viscosity this is not true anymore. To emphasise this fact, we add a subscript '$\nu$' to the notation of the potential vorticity perturbation.
It is not difficult to construct equation for $I_\nu$ using the set of equations (\ref{SFH_sys_4}-\ref{SFH_sys_6}):
\begin{equation}
\label{I_eq}
\dot I_\nu = - R^{-1} k^2 I_\nu,
\end{equation}
where the second term in the right-hand side of equation (\ref{I_eq}), explicitly ${\rm i} (2-q) R^{-1} k^2 \hat W$, has been omitted for the following reason.
In the problem we consider, the viscosity is small, $R\gg 1$, $R_b \gg 1$. Thus, in the leading orders in $R^{-1}$ and $R_b^{-1}$ one can use an inviscid balanced solution for $\hat W$ 
(see equation (37) of \citetalias{razdoburdin-zhuravlev-2015}) which gives that $|\hat W/I_\nu|\ll 1$ 
provided that $|\tilde k_x| \gg 1$. This means that the omitted term is much less than the main term in the right-hand side of equation (\ref{I_eq}).
Similarly, in the leading orders in $R^{-1}$ and $R_b^{-1}$, it is sufficient to take the slow varying solution of inhomogeneous equation (\ref{wave_visc_u_y}) in its inviscid form, 
\begin{equation}
\label{u_y_sol}
\hat u_y = \frac{\tilde k_x}{K} I_\nu,
\end{equation}
but with $I_\nu$ decaying as described by equation (\ref{I_eq}).
This is the decay of $I_\nu$ which yields an additional factor $\exp (-2/3)$ in equation (\ref{anal_G_max_compr}) being evaluated at the $t_s=-k_x/(qk_y)$ with $t_s$ additionally equated to (\ref{t_max}). 
Correspondingly, $\hat u_x$ is given by 
\begin{equation}
\label{u_x_sol}
\hat u_x = -\frac{K+4q}{K^2+4q^2k_y^2} k_y I_\nu,
\end{equation}
see equation (35) by \citetalias{razdoburdin-zhuravlev-2015}. We see that as far as $|\tilde k_x|$ is large, the radial velocity perturbation is much less than the azimuthal one in the case of tightly wound vortices, i.e. when $|\tilde k_x| \gg k_y$, 
and the velocity perturbation divergence is small:
\begin{equation}
\label{vort_property}
{\rm i}\dot{\hat W} \sim \frac{k_y}{\tilde k_x^2} \hat u_y.
\end{equation}
The tightly wound vortical spirals are of primary interest here, since we are dealing with high $R$ and focus on $k_y \lesssim 1$ which leads to 
$t_{max}\gg 1$, see equation (\ref{t_max}). Further, since $t_{max}$ approximately equals to $t_s$ of an optimal SFH exhibiting $G_{max}$, at the initial moment $|k_x| \gg k_y$ and the same is about $|\tilde k_x|$ until $t>0$ is not too close to the instant
of swing.

Equation (\ref{vort_property}) implies that terms in the second square brackets in the right-hand sides of equations (\ref{wave_visc_u_x}) and (\ref{wave_visc_u_y}) are smaller at least by a factor $|\tilde k_x|^{-3}$ than the first and the second terms therein. Taking into account that terms proportional to $R_b^{-1}$ are absent in equation (\ref{I_eq}), we come to the conclusion that bulk viscosity cannot affect the dynamics of vortical SFH even if $R_b\sim 1$.

The opposite situation takes place with density waves. These are rapidly oscillating solutions to the homogeneous equations (27-29) of \citetalias{razdoburdin-zhuravlev-2015}.
That is why, as long as $|\tilde k_x|\gg 1$, $\dot{\hat u}_x \sim |\tilde k_x| \hat u_x$ and so is about $\dot{\hat u}_y \sim |\tilde k_x| \hat u_y$ in the inviscid case.
We see that in this situation at least the term $R_b^{-1} \tilde k_x^2 \dot{\hat u}_x$ becomes the leading one among all viscous terms in equation (\ref{wave_visc_u_x}) 
and it exceeds the inviscid terms as soon as $R_b \lesssim |\tilde k_x|^2$. 

\section{Supplementary material to the global problem}
\subsection{Viscous terms}
\label{visc_app}

Viscous terms entering the dynamical equations (\ref{direct1}-\ref{direct3}) explicitly are

\begin{equation}
\label{direct4}
N_r=\frac{\nu}{r}\frac{\partial}{\partial r}\left(r\frac{\partial \delta v_r}{\partial r}\right)-\frac{\nu\left(m^2-1\right)}{r^2}\delta v_r+\frac{2}{\Sigma r}\frac{\partial\delta v_r}{\partial r}\frac{d}{dr}\left(\nu \Sigma r\right),
\end{equation}

\begin{equation}
\label{direct5}
\begin{aligned}
&N_{\varphi}=-\frac{\nu r\Omega^{\prime}}{\Sigma a_*^2}r^{-\frac{2n}{n+1}}\frac{d}{dr}\left(\Sigma r^{\frac{2n}{n+1}}\right)\delta h+\\
&+\frac{1}{\Sigma r^2}\frac{\partial}{\partial r}\left(\nu \Sigma r^3 \frac{\partial}{\partial r}\frac{\delta v_{\varphi}}{r}\right)+\frac{{\rm i}m}{\Sigma r^3}\frac{d}{d r}\left(\nu \Sigma r^2\right)\delta v_r-\frac{\nu m^2}{r^2}\delta v_{\varphi},
\end{aligned}
\end{equation}

\begin{equation}
\label{direct7}
B_r=\frac{1}{\Sigma}\frac{\partial }{\partial r}\left(\Sigma\left(\nu_b+\frac{\nu}{3}\right){\rm div}~\delta \mathbf{v}\right)-\frac{1}{\Sigma r^2}\frac{d}{d r}\left(\nu \Sigma r^2\right) {\rm div}~\delta \mathbf{v},
\end{equation}

\begin{equation}
\label{direct8}
B_{\varphi}=\frac{{\rm i}m}{r}\left(\nu_b+\frac{\nu}{3}\right){\rm div}~\delta \mathbf{v}.
\end{equation}

The norm (\ref{glob_norm}) yields the following viscous terms in the adjoint equations (\ref{adjoint1}-\ref{adjoint3}):

\begin{equation}
\label{adjoint4}
\tilde{N_r}=
-\frac{\nu}{r}\frac{\partial}{\partial r}\left(r\frac{\partial\delta\tilde{v}_r}{\partial r}\right)
+\frac{\nu\left(m^2-1\right)}{r^2}\delta \tilde{v}_r
-\frac{2}{\Sigma r}\frac{\partial \delta \tilde{v}_r}{\partial r}\frac{d}{dr}\left(\nu \Sigma r\right)
\end{equation}

\begin{equation}
\label{adjoint5}
\tilde{N}_{\varphi}=-\frac{1}{\Sigma r^2}\frac{\partial}{\partial r}\left(\nu \Sigma r^3 \frac{\partial}{\partial r}\frac{\delta \tilde{v}_{\varphi}}{r}\right)-\frac{{\rm i}m}{\Sigma r^3}\frac{d}{dr}\left(\nu \Sigma r^2\right)\delta \tilde{v}_r+\frac{\nu m^2}{r^2}\delta \tilde{v}_{\varphi}
\end{equation}

\begin{equation}
\label{adjoint6}
\tilde{N}_h=\frac{\nu r \Omega^{\prime}}{\Sigma}r^{-\frac{2n}{n+1}}\frac{d}{dr}\left(\Sigma r^{\frac{2n}{n+1}}\right) \delta \tilde{v}_{\varphi}
\end{equation}

\begin{equation}
\label{adjoint7}
\tilde{B}_r=-\frac{1}{\Sigma}\frac{\partial}{\partial r}\left(\Sigma \left(\nu_b+\frac{\nu}{3}\right) {\rm div}~\delta\tilde{\mathbf{v}}\right)
+\frac{1}{\Sigma r^2}\frac{d}{dr}\left(\nu \Sigma r^2\right) {\rm div}~\delta\tilde{\mathbf{v}}
\end{equation}

\begin{equation}
\label{adjoint8}
\tilde{B}_{\varphi}=-\frac{{\rm i}m}{r}\left(\nu_b+\frac{\nu}{3}\right) {\rm div}~\delta\tilde{\mathbf{v}}
\end{equation}
Note that boundary terms that emerge in the course of derivation of equations (\ref{adjoint1}-\ref{adjoint3}) are used to formulate 
boundary conditions for the adjoint variables, see the Appendix \ref{boundary_app}.
Also, similar to what we have in the local problem, the terms in equations (\ref{adjoint4}), (\ref{adjoint5}), (\ref{adjoint7}) and (\ref{adjoint8}) 
acquire opposite signs comparing to corresponding terms in equations (\ref{direct4}-\ref{direct8}).
Additionally, the term reciprocal to term $\propto \delta h$ in equation (\ref{direct5}) appears in the adjoint equation for $\delta \tilde h$, see equation (\ref{adjoint3}). 
Note that the last term vanishes as one turns to local perturbations.

\subsection{Boundary conditions}
\label{boundary_app}

Let us impose the boundary conditions for perturbations necessary to advance equations (\ref{direct1}) -- (\ref{direct3}) forward in time and 
equations (\ref{adjoint1}) -- (\ref{adjoint3}) backward in time. As the disc is considered to be radially infinite, only a condition at the inner boundary $r=r_i$ is relevant for the dynamics at a finite timespan.

\subsubsection{Models N1 and P1}
\label{N1_P1}

At first, let us consider the set of equations for projections of the displacement vector $\mathbf{\xi}$.
It can be done using a general relationship between $\mathbf{\xi}$ and the velocity Lagrangian perturbation \cite[Eq. 16] {lynden-bell-ostriker-1967}:
\begin{equation}
\frac{\partial \xi_r}{\partial t}+{\rm i}m\Omega\xi_r=\delta v_r+\frac{\partial}{\partial r}\left(v_r \xi_r\right)
\end{equation}

\begin{equation}
\frac{\partial \xi_{\varphi}}{\partial t}+{\rm i}m\Omega\xi_{\varphi}=\delta v_{\varphi}+r\Omega^{\prime}\xi_r+r v_r \frac{\partial}{\partial r}\left(\frac{\xi_{\varphi}}{r}\right)
\end{equation}

Since $v_r$ tends to infinity at $r_i$ (see equation (\ref{v_r2})), $\xi_r$ and $\xi_{\varphi}$ must vanish at $r_i$.
This implies that the boundary remains unperturbed, which simplifies the boundary conditions.
Now, as $v_r \propto D^{-3}$ in the vicinity of the boundary, the components of the displacement vector must behave at least as
\begin{equation}
\label{xi_r_prop}
\xi_r \propto D^3,
\end{equation}
\begin{equation}
\label{xi_phi_prop}
\xi_{\varphi} \propto D^4
\end{equation}
close to $r=r_i$.

Boundary conditions to the set (\ref{direct1}) -- (\ref{direct3}) must be that the Lagrangian perturbation of surface force acting on the inner boundary of the flow 
is equal to zero.
A relation between this force and the full stress tensor, $\sigma_{ik}$, can be found, e.g., in \citet[\S 15]{landau-lifshitz-1987}.
Since $\xi_r=0$, the Lagrangian perturbation of the normal to the boundary is equal to zero.
Thus, the boundary condition turns into two independent conditions on the Lagrangian perturbation of $\sigma_{ik}$: 
\begin{equation}
\label{sigm_rr}
\Delta \sigma_{rr}=0,
\end{equation}
\begin{equation}
\label{sigm_rph}
\Delta \sigma_{r\varphi}=0.
\end{equation}

Integration of equation (\ref{sigm_rph}) over $z$ yields
\begin{equation}
\begin{aligned}
\label{boundary1}
&\Sigma \nu \left[r\frac{\partial}{\partial r}\left(\frac{\delta v_{\varphi}}{r}\right)+\frac{{\rm i}m}{r}\delta v_r\right]+
\xi_r \frac{d}{dr}\left(\Sigma\nu r \Omega^{\prime}\right)-\\
&-2\xi_r\frac{dH}{dr}\nu r \Omega^{\prime} \rho \big{|}_{z=H}=0
\end{aligned}
\end{equation}
In equation (\ref{boundary1}) all variables are taken at the inner boundary of disc.
However, because of the relation (\ref{v_r3_integrated}) and limitations on the displacement vector (\ref{xi_r_prop}), (\ref{xi_phi_prop}) 
equation (\ref{boundary1}) is satisfied for arbitrary $\delta v_r$ and $\delta v_{\varphi}$. Thus, equation (\ref{boundary1}) does not provide us with
any restriction on perturbations at $r_i$.

At the same time, equation (\ref{sigm_rr}) results in the following equality:
\begin{equation}
\begin{aligned}
\label{boundary2}
&2\rho \nu \frac{\partial \delta v_r}{\partial r}+\rho\left(\nu_{b}-\frac{2}{3}\nu\right)\left(\frac{1}{r}\frac{\partial}{\partial r}\left(r \delta v_r\right)+\frac{{\rm i}m}{r}\delta v_{\varphi}\right)-\\
&-2\rho \nu \Omega^{\prime} \xi_{\varphi}-\delta p -\frac{\partial p}{\partial r}\xi_r=0
\end{aligned}
\end{equation}
In equation (\ref{boundary2}) all variables are taken at the inner boundary of disc.
Equation (\ref{boundary2}) allows us to formulate the necessary restrictions on $\delta v_r$ and $\delta v_{\varphi}$ at $r_i$, 
provided that it is supplied by the regularity condition for $\delta h=\delta p / \rho$ at the boundary. 
Indeed, equation (\ref{boundary2}) yields
\begin{equation}
\begin{aligned}
\label{boundary3}
&\delta h=
\frac{\delta p}{\rho}=
\left(\nu_b+\frac{4}{3}\nu\right)\left(\frac{\nu_b-\frac{2}{3}\nu}{\nu_b+\frac{4}{3}\nu}\frac{\delta v_r}{r}+\frac{\partial \delta v_r}{\partial r}\right)+\\
&+\left(\nu_b-\frac{2}{3}\nu\right)\frac{{\rm i}m}{r}\delta v_{\varphi}-
2\nu \Omega^{\prime} \xi_{\varphi}-
\frac{\partial p}{\partial r}\frac{\xi_r}{\rho}.
\end{aligned}
\end{equation}
The last two terms vanish due to limitations (\ref{xi_r_prop}), (\ref{xi_phi_prop}).
Since $\nu$ diverges as $r\to r_i$, $\delta h$ is regular at $r_i$ as soon as
\begin{equation}
\label{boundary4}
-\frac{1}{2}\frac{\delta v_r}{r}+\frac{\partial \delta v_r}{\partial r}=0,
\end{equation}
\begin{equation}
\label{boundary5}
\delta v_{\varphi}=0.
\end{equation}

As it is mentioned in \S \ref{adj_eqs}, in order to get the boundary conditions for the adjoint variables, it is necessary to put the boundary terms emerging 
in derivation of ${\bf A}^\dag$ equal to zero for each adjoint equation independently. This results in the following set of the adjoint boundary conditions:
\begin{equation}
\label{boundary_adj1}
\delta\tilde{v}_r=0
\end{equation}

\begin{equation}
\label{boundary_adj2}
\delta \tilde{v}_{\varphi}=0
\end{equation}

\begin{equation}
\label{boundary_adj3}
\delta \tilde{h}=0
\end{equation}

\subsubsection{Models N2 and P2}
\label{N2_P2}

In this case we impose the no-slip boundary conditions at $r=r_i$:
\begin{equation}
\label{boundary_simple1}
\delta v_r=0,
\end{equation}

\begin{equation}
\label{boundary_simple2}
\delta v_{\varphi}=0.
\end{equation}

For a disc of uniform surface density the substitutions in derivation of the adjoint equations vanish under the conditions
\begin{equation}
\label{boundary_adj_simple1}
\delta\tilde{v}_r=0,
\end{equation}

\begin{equation}
\label{boundary_adj_simple2}
\delta \tilde{v}_{\varphi}=0
\end{equation}
at $r=r_i$.
Note that for models N2 and P2 the boundary condition onto $\delta \tilde{h}$ is not required, since the advective terms are absent in this case, see explanation in \S \ref{num_meth}.

\end{document}